\documentclass[twocolumn,aps,prb,showpacs]{revtex4-1}
\usepackage[utf8]{inputenc}
\usepackage[normalem]{ulem}
\usepackage[T1]{fontenc}
\usepackage[unicode=true, pdfusetitle, bookmarks=true, bookmarksnumbered=false, bookmarksopen=false, breaklinks=false, pdfborder={0 0 1}, backref=false, colorlinks=false]{hyperref}
\hypersetup{colorlinks,linkcolor=blue,citecolor=blue,urlcolor=blue}
\usepackage{float}
\usepackage{verbatim}
\usepackage[abs]{overpic}
\usepackage{color}
\usepackage{comment}
\usepackage{amsmath,amssymb}
\usepackage{graphicx}

\begin{document}

\title{Formation probabilities and statistics of observables as defect problems in free fermions and quantum spin chains}

\author{M. N. Najafi}
\affiliation{Department of Physics, University of Mohaghegh Ardabili, P.O. Box 179, Ardabil, Iran}
\author{ M.~A.~Rajabpour}
\affiliation{  Instituto de F\'isica, Universidade Federal Fluminense, Av. Gal. Milton Tavares de Souza s/n, Gragoat\'a, 24210-346, Niter\'oi, RJ, Brazil}
\date{\today{}}

\begin{abstract}
We show that the computation of formation probabilities (FP's) in the configuration basis and the full counting statistics of observables in quadratic fermionic Hamiltonians are equivalent to the calculation of emptiness formation probability (EFP) in the Hamiltonian with a defect. In particular, we first show that the FP  of finding a particular configuration in the ground state is equivalent to the EFP of the ground state of the quadratic Hamiltonian with a defect. Then, we show that the probability of finding a particular value for any quadratic observable is equivalent to a FP problem and ultimately leads to the calculation of EFP in the ground state of a Hamiltonian with a defect. We provide exact determinant formulas for the FP in generic quadratic fermionic Hamiltonians. For applications of our formalism we study the statistics of the number of particles and kinks. Our conclusions
can be extended also to quantum spin chains, which can be mapped to free fermions via Jordan-Wigner transformtion. In particular, we 
provide an exact solution to the problem of the transverse field $XY$ chain with a staggered line defect. We also study the distribution of magnetization and kinks in the transverse field $XY$ chain numerically and show how the dual nature of these quantities manifests itself in the distributions.
\end{abstract}

\maketitle

\section{Introduction}
\label{sec:intro}
Consider a quantum many-body state written in a configuration basis, then the formation probability (FP) of a particular configuration in a subsystem is the probability of finding the configuration if we do the measurement in that particular basis.
For example, if we take the ground state of a  spinless fermionic system and ask the same question for the subsystem $D$ then there are $2^{|D|}$ possibilities for the configurations, where $|D|$ is the size of the subsystem. Every configuration appears with a particular probability which we call formation probability \cite{Najafi2016}. The simplest example of the FP is the emptiness formation probability (EFP) which is about the FP of the configuration without any fermion in the subsystem and has a long history. It was studied in the context of the $XXZ$ spin chain in \cite{Korepin1994,Korepin1995,Kitanine2002a,Kitanine2002b,Korepin2002,Korepin2003,Cantini2012} and in the context of the $XY$ chain in \cite{Shiroishi2001,Franchini2003,Franchini2005,Viti2019}.
FP has been also studied from the conformal field theory (CFT) point of view in \cite{Stephan2014,Rajabpour2015,Rajabpour2016,viti2016}. The problem is related to the solution of the free energy of a quantum field theory with a slit in the context of the CFT studied in \cite{Dubail2013} and references therein. A relation to the Casimir energy problem is also established in \cite{Rajabpour2015,Rajabpour2016}. The FP can be also used to find the Shannon entropy with a plethora of applications in the studies of quantum phase transitions 
(see~\cite{Stephan2009,AR2013,Stephan2014PRB,AR2014,Luitz2014a,Luitz2014b,AR2015,Alcaraz2016,NR2014,Alba2016,NR2019}).

One can also look to the problem of the FP from a different point of view. Consider a configuration basis which is associated
with the local on-site observable $\hat{o}_i$ at site $i$ with eigenvalues(eigenvectors) $o_i^j$($|o_i^j\rangle$), $j=1,2,...,d$, where $d$ is the dimension of the local on-site Hilbert space. Then 
one can define an on-site
projection operator as $\pi_i^j=|o_i^j\rangle\langle o_i^j|$. Multiplication of these operators in a subsystem leads to a projection operator $\Pi_{\hat{o}}^{\{j\}}=\prod_i\pi_i$, where 
the set ${\{j\}}$ fixes the configuration by picking a particular $\pi^j$ at every site. Finally the  
average over, for example, the ground state $|g\rangle$ of the Hamiltonian $H$ gives the FP
for the desired configuration, i.e. $p\left({\{j\}}\right)=\langle g|\Pi_{\hat{o}}^{\{j\}}|g\rangle$. Now consider one {\it{primitive}} configuration with all the sites being in the eigenstate corresponding
to, for example, $o^1$
i.e. $|\{1\}\rangle$. In the case of fermions this primitive configuration can be 
the configuration without fermions. Then one can always find a unitary similarity transformation matrix $T^{\{j\}}$ which $p({\{j\}})=\langle g'|\Pi_{\hat{o}}^{\{1\}}|g'\rangle$, where $|g'\rangle=T^{\{j\}}|g\rangle$.
The state $|g'\rangle$ can be considered as the ground state of the Hamiltonian $H'=T^{\{j\}}H\left(T^{\{j\}}\right)^{-1}$ which is basically the same Hamiltonian as $H$ but with a line defect.
This simple argument shows that formally one can look to  the problem of generic FP as 
the problem of the probability of a primitive configuration in the deformed Hamiltonian.
Many-body systems with line defects have been studied for decades; for some earlier studies 
see, for example, \cite{PFEUTY1979245,Bariev1979,McCoy1980,Kadanov1981,Brown1982,Cabrera1987,Abraham1989,Hinrichsen1989}. For studies related to boundary CFT and integrable quantum field theories
see \cite{Turban1985,Guimaraes1986,Henkel1987,Henkel1988,Grimm1989,Zhang1996,Oshikawa1997} and \cite{Mussardo1994}.

In a different approach to the above one could also define the generating function $M(\{\lambda_i\})=\langle e^{\sum_i\lambda_i\hat{o}_i}\rangle$, where the coefficients of the 
exponentials are the FP's. This is a nontrivial example of a more widely known concept called full counting statistics (FCS) 
which deals with the fluctuations of an observable defined in a subsystem. In most of the FCS problems one is interested in the moments of an observable $\hat{\mathcal{O}}$ defined for the entire subsystem, i.e. $\langle\hat{\mathcal{O}}^n\rangle$. 
Being a very natural concept FCS has been studied for a long time in different communities.
It has been studied in the context of charge fluctuations \cite{Levitov1993,Klich2014}, Bose gases \cite{Gritsev2006,Hofferberth2008,Bastianello2018,Piroli2018,vanNieuwkerk2018}, 
particle number fluctuations \cite{Song2012,Vicari2012,Rachel2012,Herviou2017,Herviou2019}, quantum spin chains \cite{Eisler2003,Demler2007,Fendley2008,Ivanov2013,De-Chiara2016,NR2017,Collura2017,Groha2018}, and out of equilibrium quantum systems \cite{Essler2018,vanNieuwkerk2018,Collura2019}.

It is not difficult to see that the FCS of an observable defined in a subsystem can be formulated as a formation probability problem by just 
considering the basis that diagonalizes the observable $\hat{\mathcal{O}}$. In this basis one can consider every eigenstate of the observable as a {\it{configuration}} and then 
the probability to find a particular value for the observable is just the FP for that {\it{configuration}}. This connection seems too formal to be useful; however, in this paper we show that this point of view is extremely useful when one deals with quadratic observables in the study of FCS in the quadratic fermion Hamiltonians and the corresponding spin chains.
The main advantage with respect to the more standard generating function point of view is that here one does not need to do the inverse Laplace transformation which is normally
impossible to do analytically. Even numerical calculation of such kind of inverse transformation is usually extremely difficult because of the presence of a square root of a determinant
in the final result of the generating function. Apart from its conceptual 
appeal our approach provides an explicit formula for the probabilities which is its main advantage with respect to the generating function method.

\subsection{Organization of the paper and summary of the results}
\label{subsec:summary}
The result of this paper can be summarized in three different parts. In the first part, i.e., Sec. \ref{sec:2}, 
we first introduce the concept of FP and EFP in the quadratic Hamiltonians. Then we provide new determinant formulas for the 
FP and show explicitly how the FP is an EFP for the Hamiltonian with a defect, a line defect in the case of one-dimensional systems. The main equations of this section are (\ref{main relations a}) and (\ref{main relations b}), which give explicit formulas for the formation probability of the ground state of a quadratic free fermion Hamiltonian. The results of this section are independent of the dimension and can be also used for the states in which Wick's theorem is valid.
For example, one can use the same formulas to calculate the formation probabilities for the excited states too. It is worth mentioning that since we have determinant formulas for the formation probabilities the complexity of calculating them increases polynomially with the size of the subsystem. Clearly due to the exponential number of probabilities calculating all of them is an exponential problem. 

In the second part, i.e. Sec. \ref{sec:3}, we study the statistics of quadratic observables and show that this problem is related to the FP and ultimately EFP of a Hamiltonian with a defect. Here we give an explicit formula, i.e., (\ref{generic-probability}), for the probability of finding a particular value for a generic quadratic observable when the total system is in the ground state. Here too all the results are independent of the dimension and can be generalized to the states in which Wick's theorem is valid, such as the excited states. In this section a couple of examples including the number of particles and kinks will be discussed briefly. Similar to the calculation of the FP if one is interested to calculate the probability of finding one particular value for the observable our formulas provide a method to calculate it in a time which grows polynomially with the size of the subsystem. Calculating the full distribution is in general an exponentially time consuming problem.

The last part of this paper, i.e. Sec. \ref{sec:4} and \ref{sec:5}
are about the explicit application of the formalisms of the previous sections.
We first study the transverse field $XY$ chain with a staggered line defect. We find the exact correlation matrix of the ground state of this Hamiltonian which gives the probability of the Neel formation, i.e. (\ref{Eq:staggeredMain1}) in the original Hamiltonian without any defect. Then we argue that our result is valid for a generic translational invariant
quadratic fermion Hamiltonian with a staggered line defect. 
Then in Sec. \ref{sec:5} we use the general equations introduced in Sec. III to calculate numerically the probability distribution of the particle numbers and kinks in the transverse field $XY$ chain.
Finally in Sec. VI we summarize our findings and comment on the future directions.

\section{Formation probability as an emptiness formation probability}
\label{sec:2}
Consider the following free fermion Hamiltonian with real generic couplings: 
\begin{equation}\label{H1}\
\mathcal{H}_{\text{free}}(\textbf{A},\textbf{B})=\textbf{c}^{\dagger}\textbf{A}\textbf{c}+\frac{1}{2}\textbf{c}^{\dagger}\textbf{B}\textbf{c}^{\dagger}+\frac{1}{2}\textbf{c}\textbf{B}^{T}\textbf{c}-\frac{1}{2}{\rm Tr}{\textbf{A}},
\end{equation} 
where $\textbf{A}$ and $\textbf{B}$ are symmetric and anti-symmetric matrices respectively and $\textbf{c} \equiv (c_1,c_2,...,c_{|D|})$  with similar 
definition for $\textbf{c}^{\dagger}$. We define the correlation matrix of the ground state of the above Hamiltonian as:
\begin{eqnarray}\label{G normal}\
iG_{jk}&=&\langle \bar{\gamma}_j\gamma_k\rangle,
\end{eqnarray} 
where we defined the Majorana operators $\gamma_k=c_k+c_k^{\dagger}$ and $\bar{\gamma}_j=i(c_j^{\dagger}-c_j)$ and $\left\langle \right\rangle $ is normally
the expectation value in the ground state. Note that here we have
$\delta_{jk}=\langle \gamma_j\gamma_k\rangle=-\langle \bar{\gamma}_j\bar{\gamma_k}\rangle$. Using Wick's theorem which is valid for the eigenstates of the above Hamiltonian one can write all the other correlation functions with respect to the three basic correlation functions, i.e., $\langle c^{\dagger}_jc_k\rangle$, $\langle c_jc_k\rangle$, and $\langle c^{\dagger}_jc^{\dagger}_k\rangle$, which can be re-written in terms of $G_{jk}$. Other interesting quantities such as entanglement entropy and FP's can be also written as a function of the correlation matrix. Intuitively FP is defined as the probability of finding a particular 
configuration $C$ in a subsystem of the full system. It was shown in \cite{Najafi2016} that the result can be written with respect to the correlation matrix as follows:
\begin{eqnarray}\label{Main NR2015}\
p(C)=\det\frac{\mathbf{I}-\mathbf{G}}{2}\text{Min}[\mathbf{F}],
\end{eqnarray} 
where $\text{Min}\mathbf{F}$ is a particular principal minor of the matrix $\mathbf{F}=\frac{\mathbf{I}+\mathbf{G}}{\mathbf{I}-\mathbf{G}}$ derived after
removing the rows and columns of the sites without any fermion in the configuration $C$. When there is no fermion in the configuration $C$ then the corresponding probability is called
 emptiness formation probability and we have
\begin{eqnarray}\label{emptiness}\
p(\{0\})=\det\frac{\mathbf{I}-\mathbf{G}}{2},
\end{eqnarray} 
 whereas when all the sites are occupied with fermions we have
\begin{eqnarray}\label{fulliness}\
p(\{1\})=\det\frac{\mathbf{I}+\mathbf{G}}{2}.
\end{eqnarray} 
The above two formulas are very useful equations to do analytical calculations when there is a translational invariance. However,
because of the presence of the minor in Eq. (\ref{Main NR2015}) analytical calculations do not seem to be feasible in more generic cases. One way to overcome this problem is to map the problem of FP to the problem of EFP and see the outcome. This procedure
can be done as follows: Consider the formation probability, i.e. $P(C)$, of the ground state of the Hamiltonian $\mathcal{H}_{\text{free}}(\textbf{A},\textbf{B})$.
This probability is equal to the EFP of the ground state of the Hamiltonian $\mathcal{H}_{\text{free}}(\textbf{A}',\textbf{B}')$
with a defect. In Sec. IV we show for an explicit example how this procedure can be followed. Finally we have:
\begin{eqnarray}\label{formation-emptiness}\
p(C)=\det\frac{\mathbf{I}-\mathbf{G'}}{2},
\end{eqnarray} 
where $\mathbf{G'}$ is the correlation matrix of the Hamiltonian with defects.
Following \cite{LIEB1961407} one can, in principle, find this correlation matrix numerically for any values of the coupling constants. However, analytical calculations are 
commonly feasible when
we have translational invariance or some extra structure. For example, the Hamiltonian of a translationally invariant (periodic) free fermion with time-reversal symmetry can be written as
\begin{equation}\label{Hamiltonian}\
	H=\sum_{r=-R}^R\sum_{j\in\Lambda}a_rc_j^{\dagger}c_{j+r}+\frac{b_r}{2}(c_j^{\dagger}c_{j+r}^{\dagger}-c_jc_{j+r})+\text{const}.\hspace{0.33cm}
\end{equation} 
Using the Majorana operators one can also write,
\begin{eqnarray}\label{Hamiltonian-majorana}\
H=\frac{i}{2}\sum_{r=-R}^R\sum_{j\in\Lambda}t_r\bar{\gamma}_j\gamma_{j+r};
\end{eqnarray} 
where $t_r=-(a_r+b_r)$ and $t_{-r}=-(a_r-b_r)$. It is very useful to put the coupling constants as the coefficients of the following holomorphic function $f(z)=\sum_rt_rz^r$. 
Then the Hamiltonian can be diagonalized by going to the Fourier space and then Bogoliubov transformation as follows (see, for example,~\cite{Jones2019}):
\begin{eqnarray}\label{Hamiltonian-diagonalization}\
	H=\sum_q|f(q)|\eta_q^{\dagger}\eta_q+\text{const},
\end{eqnarray} 
where $\eta_q=\frac{1}{2}(1+\frac{f(q)}{|f(q)|})c_q^{\dagger}+\frac{1}{2}(1-\frac{f(q)}{|f(q)|})c_{-q}$ with $f(q):=f(e^{iq})$. 
Finally, in the thermodynamic limit ($L\rightarrow\infty$, transforming the summation over discrete $q$ values to the integral over $q$) one can write the following explicit formula for the correlation matrix of the ground state:
\begin{eqnarray}\label{G matrix translational invariance}\
G_{jk}=\int_{0}^{2\pi}\frac{\text{d}q}{2\pi}\frac{f(q)}{|f(q)|}e^{iq(j-k)}.
\label{Eq:G}
\end{eqnarray} 
The above correlation matrix has a Toeplitz structure which makes it a suitable candidate for analytical calculations.
It is possible to extend the above result to also excited states without much difficulty. When there is no translational invariance as it is the case 
for $\mathcal{H}_{\text{free}}(\textbf{A}',\textbf{B}')$
following the above procedure is not simple.
An explicit calculation will be presented later for the transverse field $XY$ chain with the staggered magnetization. 
In this paper we will follow another path which is going to be one of our main results. 

The basic idea of the second method is based on writing Eq.~ (\ref{Main NR2015}) like Eq.~ (\ref{formation-emptiness}). We show here that there are at least two different ways
to do this. Consider a generic principal minor of a generic matrix $\mathbf{M}$. The basic idea is to write the principal minor as follows:
\begin{eqnarray}\label{main relations0}\
\text{Min}[\mathbf{M}]&=&\det\frac{\mathbf{M}(\mathbf{I}-\mathbf{I}_c)+\mathbf{I}+\mathbf{I}_c}{2};\\
\text{Min}[\mathbf{M}]&=&\det\frac{(\mathbf{I}-\mathbf{I}_c)\mathbf{M}+\mathbf{I}+\mathbf{I}_c}{2};
\end{eqnarray}
where $\mathbf{I}$ is an identity matrix and $\mathbf{I}_c$ is a diagonal matrix made out of $\pm1$ which clearly depends on which columns and rows are getting removed. We
set its diagonal element to $-1$ when we have a fermion and $1$ when there is no fermion at the corresponding
site. Now it is easy to show that we have
\begin{eqnarray}\label{main relations a}\
p(C)&=&\det\frac{\mathbf{I}-\mathbf{G}\mathbf{I}_c}{2},\\
\label{main relations b}
p(C)&=&\det\frac{\mathbf{I}-\mathbf{I}_c\mathbf{G}}{2}.
\end{eqnarray}
The above equations mean that the formation probability $p(C)$ is actually the  EFP for a defect Hamiltonian 
with the correlation matrix $\mathbf{G}\mathbf{I}_c$ or $\mathbf{I}_c\mathbf{G}$. As we mentioned before  none of these correlation matrices are necessarily the actual correlation matrix
of the defect Hamiltonian $\mathcal{H}_{\text{free}}(\textbf{A}',\textbf{B}')$ introduced above. In fact we will show explicitly later that for the staggered Ising chain the 
correlation matrix has quite a different form. This should not be surprising because one can extract the minor of a matrix using quite different methods and although they
all end up with the same number for $p(C)$ they have been derived from different matrices. However, clearly
finding one is enough to get the others by proper manipulation of the rows and columns of the correlation matrix. For example, if $\mathbf{G}'$ is the actual correlation matrix then
there is a similarity transformation $\mathbf{S}$ and we have $\mathbf{G}'=\mathbf{S}^{-1}\mathbf{G}\mathbf{I}_c\mathbf{S}$. For generic correlation matrices finding the $\mathbf{S}$
matrix 
is not necessarily an easy problem. It is worth mentioning that using the correlation matrices $\mathbf{G}\mathbf{I}_c$ or $\mathbf{I}_c\mathbf{G}$ leads to the same set of
 FP's
as the $\mathbf{G}$ matrix.
In principle, Eqs.~ (\ref{main relations a}) and (\ref{main relations b}) probably can be useful for analytical
calculations when the $\mathbf{G}$ matrix is a Toeplitz matrix and the configuration $C$ has a pattern. In these cases the $\mathbf{G}\mathbf{I}_c$
has always a block Toeplitz structure. As an explicit example consider the ground state of the Hamiltonian (\ref{Hamiltonian}) and let us focus on the probability of the configuration
$C=(s_1,s_2,...,s_l)$, where $s_j=-1$ or $+1$ depending on the presence or the lack of a fermion at site $j$. Then we can write:
\begin{eqnarray}\label{defect G matrix }\
(\mathbf{G}\mathbf{I}_c)_{jk}=\text{sgn}_r(j,k)\int_{0}^{2\pi}\frac{\text{d}q}{2\pi}\frac{f(q)}{|f(q)|}e^{iq(j-k)},\\
(\mathbf{I}_c\mathbf{G})_{jk}=\text{sgn}_l(j,k)\int_{0}^{2\pi}\frac{\text{d}q}{2\pi}\frac{f(q)}{|f(q)|}e^{iq(j-k)},
\end{eqnarray} 
where the matrices $\textbf{sgn}_r$ and $\textbf{sgn}_l$ are the sign matrices and, for example, for a configuration with four sites have the following forms: 
\begin{eqnarray}\label{sing matrices}\
\begin{split}
&\textbf{sgn}_l=\left(
\begin{array}{cccc}
s_1 & s_1 & s_1 & s_1 \\
s_2 & s_2 & s_2 & s_2 \\
s_3 & s_3 & s_3 & s_3 \\
s_4 & s_4 & s_4 & s_4
\end{array}
\right),\\
&\textbf{sgn}_r=\left(
\begin{array}{cccc}
s_1 & s_2 & s_3 & s_4 \\
s_1 & s_2 & s_3 & s_4 \\
s_1 & s_2 & s_3 & s_4 \\
s_1 & s_2 & s_3 & s_4
\end{array}
\right).
\end{split}
\end{eqnarray} 
The generalization for bigger sizes is straightforward. We note that when the configuration has a crystal structure the above matrices have block Toeplitz forms.

\section{Statistics of a generic quadratic observable as a formation probability}
\label{sec:3}
In this section we argue that the problem of finding the statistics of a generic quadratic observable is essentially a  FP problem and consequently
the formulas derived in the previous section have many more applications than at first might appear. Consider the following quadratic observable
\begin{eqnarray}\label{Observable}\
\mathcal{O}_{D}=\textbf{c}^{\dagger}\textbf{M}\textbf{c}+\frac{1}{2}\textbf{c}^{\dagger}\textbf{N}\textbf{c}^{\dagger}+
\frac{1}{2}\textbf{c}\textbf{N}^{T}\textbf{c}-\frac{1}{2}{\rm Tr}{\textbf{M}},
\end{eqnarray} 
where $\textbf{M}$ and $\textbf{N}$ are symmetric and anti-symmetric matrices respectively. It is much more convenient to write the above observable in the following form:
\begin{eqnarray}\label{Observable2}\
\mathcal{O}_{D}=\frac{1}{2}(\textbf{c}^{\dagger}\,\,\textbf{c})\begin{pmatrix}
\textbf{M} & \textbf{N}\\
-\textbf{N} & -\textbf{M}\\
  \end{pmatrix}
  \begin{pmatrix}
\textbf{c}\\
\textbf{c}^{\dagger}\\
  \end{pmatrix},
\end{eqnarray}
Note that the above observable can have support just in a subsystem $D$ of the full system. With the statistics of this observable we mean the probability of finding a particular value if we measure the above quantity if the full system is in the ground state. We prove here that this is a FP problem. To show this we first 
diagonalize the observable $\mathcal{O}_{D}$ with the standard method of Ref.~\cite{LIEB1961407} (see Appendix A). The idea is based on a canonical transformation 
\begin{eqnarray}\label{mat_U1}\
\begin{pmatrix}
\textbf{c} \\
\textbf{c}^{\dagger} \\
\end{pmatrix}=
\textbf{U}^{\dagger}
\begin{pmatrix}
\boldsymbol{\delta} \\
\boldsymbol{\delta}^{\dagger} \\
\end{pmatrix},
\end{eqnarray} 
which leads to 
\begin{eqnarray}\label{Hdiag1}\
\mathcal{O}_{D}=\sum_{k}|\lambda_{k}|(\delta_{k}^{\dagger}\delta_{k}-\frac{1}{2}).
\end{eqnarray} 
The eigenvalues and the eigenvectors of the observable can be derived as usual by applying the modes on the {\it{ground-state}} properties.
The next step is to write the Hamiltonian with respect to the $\delta_{k}^{\dagger}$ and $\delta_{k}$ as follows:
\begin{equation}\label{Hamiltonian New Form}\
\mathcal{H}_{\text{free}}(\textbf{A},\textbf{B})=\frac{1}{2}(\boldsymbol{\delta}^{\dagger}\,\,\boldsymbol{\delta})\textbf{U}\begin{pmatrix}
\textbf{A} & \textbf{B}\\
-\textbf{B} & -\textbf{A}\\
  \end{pmatrix}\textbf{U}^{\dagger}
  \begin{pmatrix}
\boldsymbol{\delta}\\
\boldsymbol{\delta}^{\dagger}\\
  \end{pmatrix},
\end{equation}
The above equation can now be used to make the main argument. In the new basis if one calculates the FP with the new matrices 
the result is exactly equal to the probability of finding a particular value for the corresponding observable. For example, the  EFP
for the above Hamiltonian is exactly equal to the probability of finding the minimum value for the corresponding observable $\mathcal{O}_{D}$. To find the exact formula one first needs 
to calculate the correlation matrix 
\begin{eqnarray}\label{G updated}\
iG'_{jk}&=&\langle \bar{\alpha}_j\alpha_k\rangle,
\end{eqnarray} 
where we defined the new Majorana operators $\alpha_k=\delta_k+\delta_k^{\dagger}$ and $\bar{\alpha}_j=i(\delta_j^{\dagger}-\delta_j)$. Having found the above correlation matrix the rest
of the calculation is exactly as described in the previous section. 

When the observable is defined for a subsystem one can also use the procedure that was outlined in \cite{NR2019} which is based on the reduced density matrix. When the system is in the ground state
the reduced density matrix can be written as:
\begin{eqnarray}\label{Eq:reducedDM}
\rho_{D}&=&\det\frac{1}{2}(\mathbb{I}-\textbf{G})e^{\mathcal{H}},\\
\mathcal{H}&=&\frac{1}{2}(\textbf{c}^{\dagger}\,\,\textbf{c})\begin{pmatrix}
\textbf{P} & \textbf{Q}\\
-\textbf{Q} & -\textbf{P}\\
  \end{pmatrix}
  \begin{pmatrix}
\textbf{c}\\
\textbf{c}^{\dagger}\\
  \end{pmatrix}+\frac{1}{2}{\rm Tr}\ln{(\textbf{F}_{s})},
\end{eqnarray}
where $\mathcal{H}$ is the entanglement Hamiltonian and 
\begin{eqnarray}\label{mat_T3}\
\begin{pmatrix}
\textbf{P} & \textbf{Q}\\
\textbf{-Q} & \textbf{-P}\\
  \end{pmatrix}=
\ln \begin{pmatrix}
\textbf{F}_{s}-\textbf{F}_{a}\textbf{F}_{s}^{-1}\textbf{F}_{a} & \textbf{F}_{a}\textbf{F}_{s}^{-1}\\
-\textbf{F}_{s}^{-1}\textbf{F}_{a}& \textbf{F}_{s}^{-1}\\
  \end{pmatrix},
\end{eqnarray} 
where $\textbf{F}_{a}=\frac{\textbf{F}-\textbf{F}^T}{2}$ and $\textbf{F}_{s}=\frac{\textbf{F}+\textbf{F}^T}{2}$ and as before $\mathbf{F}=\frac{\mathbf{I}+\mathbf{G}}{\mathbf{I}-\mathbf{G}}$.
Note that the $\mathbf{G}$ matrix here is calculated for the original creation and annihilation operators appearing in the Hamiltonian. The idea is again based on writing the entanglement Hamiltonian 
in the basis that the observable is diagonal, i.e., $\delta$ basis.
Now, we introduce the fermionic coherent states, i.e., $|\boldsymbol{\gamma}\rangle= |\gamma_1,\gamma_2,...,\gamma_{|D|}\rangle= e^{-\sum_{k=1}^{|D|}\gamma_{_{k}}\delta_k^{\dagger}}|0\rangle,$
 where $\gamma_k$'s 
are Grassmann numbers with the following properties: $\gamma_n\gamma_m+\gamma_m\gamma_n=0$ and $\gamma_n^2=\gamma_m^2=0$. Here, $|D|$ 
is the number of sites in the region $D$. Then, one can write~\cite{NR2019}: 
\begin{equation}\label{rhod8}
\langle \boldsymbol{\gamma} |\rho_{D}|\boldsymbol{\gamma}'\rangle=
\det\frac{1}{2}(\mathbb{I}-\textbf{G})\left[\frac{{\rm \det}(\textbf{F}_{s})}{{\rm \det}(\tilde{\textbf{F}}_{s})}\right]^{\frac{1}{2}} \,\,e^{\frac{1}{2}(\bar{\boldsymbol{\gamma}}-\boldsymbol{\gamma'})\tilde{\textbf{F}}(\bar{\boldsymbol{\gamma}}+\boldsymbol{\gamma'})}\hspace{0.5cm}
\end{equation}
where we defined
\begin{eqnarray}\label{BB_rel3}
\tilde{\textbf{F}}_{s}=e^{\tilde{\textbf{Y}}},\hspace{1cm}
\tilde{\textbf{F}}=\tilde{\textbf{X}}+e^{\tilde{\textbf{Y}}},
\end{eqnarray}
with 
\begin{equation}\label{matrixes}\
\tilde{\textbf{X}}=\tilde{\textbf{T}}_{12}(\tilde{\textbf{T}}_{22})^{-1},\hspace{0.15cm}
\tilde{\textbf{Z}}=(\tilde{\textbf{T}}_{22}^{-1})\tilde{\textbf{T}}_{21},\hspace{0.25cm}
e^{-\tilde{\textbf{Y}}}=\tilde{\textbf{T}}_{22}^{T},
\end{equation} 
where
\begin{eqnarray}\label{Ttilde}
\begin{split}
\tilde{\textbf{T}}&=\begin{pmatrix}
\tilde{\textbf{T}}_{11} & \tilde{\textbf{T}}_{12}\\
\tilde{\textbf{T}}_{21} & \tilde{\textbf{T}}_{22}\\
\end{pmatrix}\\
&=\textbf{U}\begin{pmatrix}
\textbf{F}_{s}-\textbf{F}_{a}\textbf{F}_{s}^{-1}\textbf{F}_{a} & \textbf{F}_{a}\textbf{F}_{s}^{-1}\\
-\textbf{F}_{s}^{-1}\textbf{F}_{a}& \textbf{F}_{s}^{-1}\\
\end{pmatrix}\textbf{U}^{\dagger}.
\end{split}
\end{eqnarray}
Eq. (\ref{rhod8}) can be used to calculate all the desired probabilities using the same method that was developed in Ref.\cite{Najafi2016}.
First of all, it is easy to see that to find the probability of finding the observable in its minimum value one needs to put all the $\gamma$'s equal to zero, then we have
\begin{eqnarray}\label{probability of min}
p(o_{\text{min}})=\det\frac{1}{2}(\mathbb{I}-\textbf{G})\left[ \frac{{\rm \det}(\textbf{F}_{s})}{{\rm \det}(\tilde{\textbf{F}}_{s})}\right]^{\frac{1}{2}}.
\end{eqnarray}
To find the  probability of other values, say $o$, one needs to know the corresponding modes $\lambda_k$'s
which generate the desired eigenvalue of the observable and then, perform a Grassmann integral over the corresponding $\gamma_k$'s and put the other $\gamma$'s equal to zero. The result is
\begin{eqnarray}\label{generic-probability}
p(o)=\det\frac{1}{2}(\mathbb{I}-\textbf{G})\left[\frac{{\rm \det}(\textbf{F}_{s})}{{\rm \det}(\tilde{\textbf{F}}_{s})}\right]^{\frac{1}{2}}\sum_g \text{Min} \left[\tilde{\textbf{F}}\right],
\end{eqnarray}
where $\text{Min} \left[ \tilde{\textbf{F}}\right]$ is the corresponding principal minor of the matrix $\tilde{\textbf{F}}$ and the sum takes care of the degeneracies. Note 
that one can again use Eq. (\ref{main relations0}) to get rid of the minor in the above equation. The above equation is very convenient to get explicit results
for the probabilities without going through the generating function formalism.

\subsection{Statistics of the number of particles}
\label{sec:stat-number}
Statistics of the number of particles in a subsystem is the simplest possible example one can imagine because the observable itself is already diagonal. For earlier 
detailed studies regarding fluctuations of the particles from the generating function point of view see~\cite{Song2012,Vicari2012,Rachel2012}.
It is simple to see that the probability of finding no
particle in the subsystem of size $l$ is exactly the EFP. Then the probability of finding one particle is just about summing over all the FP's with 
just one fermion. In other words we need to first calculate the sum of the minors of rank $1$ of the matrix $\mathbf{F}$. For generic case we need to find the sum of the minors of a particular rank
of the matrix. There is a standard method to calculate these numbers which is called the Faddeev-LeVerrier algorithm~\cite{leverrier1840variations} (see also Wikipedia). The probability of having $n$ particles can be explicitly written as:
\begin{eqnarray}\label{probability of numbers}
p(n)=\det\left[ \frac{1}{2}(1-\boldsymbol{G})\right](-1)^nc_{l-n},
\end{eqnarray}
where the exact form of the coefficients can be written as:
\begin{eqnarray}\label{coefficiants Bell polynomial}
\begin{split}
c_{l-n}=(-1)^{l-n}\frac{1}{n!}B_n(&\text{tr}\mathbf{F},-\text{tr}\mathbf{F}^2,2!\text{tr}\mathbf{F}^3,...\\
&...,(-1)^{n-1}(n-1)!\text{tr}\mathbf{F}^n),
\end{split}
\end{eqnarray}
where $B_n$ is the complete  exponential Bell polynomial~\cite{leverrier1840variations,bell1934exponential} (see also Wikipedia). The complete  exponential Bell polynomial can be written as:
\begin{equation}\label{complete Bell polynomial}\
B_n(x_1,x_2,...,x_{n-k+1})=\sum_{k=1}^nB_{n,k}(x_1,x_2,...,x_{n-k+1}),
\end{equation}
where the partial exponential Bell polynomial $B_{n,k}$ is given by the equation.
\begin{widetext}
\begin{eqnarray}\label{Bell polynomial0}
B_{n,k}(x_1,x_2,...,x_{n-k+1})=\sum\frac{n!}{j_1!j_2!...j_{n-k+1}!}\left(\frac{x_1}{1!}\right)^{j_1}\left(\frac{x_2}{2!}\right)^{j_2}...\left(\frac{x_{n-k+1}}{(n-k+1)!}\right)^{j_{n-k+1}},
\end{eqnarray} 
\end{widetext}
where the sum is over all the non-negative $j_1,j_2,...,j_{n-k+1}$ in a way that we have $j_1+j_2+...+j_{n-k+1}=k$ and $j_1+2j_2+...+(n-k+1)j_{n-k+1}=n$.

The first few terms of the coefficients can be written as:
\begin{eqnarray}\label{coefficiants Bell polynomial examples}
c_{l}&=&1,\\
c_{l-1}&=&-\text{tr}\mathbf{F},\\
c_{l-2}&=&\frac{1}{2!}((\text{tr}\mathbf{F})^2-\text{tr}\mathbf{F}^2),\\
c_{l-3}&=&-\frac{1}{3!}((\text{tr}\mathbf{F})^3-3\text{tr}\mathbf{F}^2\text{tr}\mathbf{F}+2\text{tr}\mathbf{F}^3).
\end{eqnarray}
Note that we also have $c_0=(-1)^l\det \mathbf{F}$. It is worth mentioning that using the properties of Bell's polynomial one can also write the following recursion
relation for the probabilities:
\begin{eqnarray}\label{probabilities recursion}
p(n)=-\frac{1}{n}\sum_{j=1}^n(-1)^j\text{tr}\mathbf{F}^jP(n-j).
\end{eqnarray}
The above formulas indicate that the problem of finding statistics of the number of particles boils down to the calculation of the trace of different powers of the matrix $\mathbf{F}$.

\subsection{Statistics of the kinks}\label{SEC:kinks}

In this section we provide an example to show how the formalism of this section should be applied for a nontrivial observable. We would like to study the statistics
of the following quantity:
\begin{eqnarray}\label{Kink}
\mathcal{K}=\frac{l-1}{2}+\frac{1}{2}\sum_{j=1}^{l-1}(c_j^{\dagger}-c_j)(c_{j+1}^{\dagger}+c_{j+1})
\label{Eq:kink}
\end{eqnarray}
The reason that we call this quantity kink statistics comes from the spin representation of this quantity after Jordan-Wigner (JW) transformation, i.e., $c_j^{\dagger}=\prod_{l=1}^{j-1}\sigma_l^z\sigma_j^{+}$,
which leads to $\mathcal{K}=\frac{1}{2}\sum_{j=1}^{l-1}(1-\sigma_j^x\sigma_{j+1}^x)$. In this example we have:
\begin{equation}\label{Kink M and N}
M_{ij}=\frac{1}{2}\delta_{i,j+1}+\frac{1}{2}\delta_{i+1,j},\hspace{0.4cm}N_{ij}=\frac{1}{2}\delta_{i+1,j}-\frac{1}{2}\delta_{i,j+1}
\end{equation}
Using the method of Appendix~\ref{app1} one can diagonalize $\mathcal{K}$. The $\textbf{C}$ matrix is simply $C_{ij}=\delta_{ij}-\delta_{i,1}\delta_{j,1}$, 
and its eigenvectors $\psi_i(j)\equiv\psi_{ij}$ are chosen to be $\delta_{i,j}$ ($i$ being the label of the eigenvector). Showing the eigenvectors by $\kappa_i$, and using Eq.~\ref{phi and psi 1a} for $\kappa_i\ne 0$, we easily show that $\phi_{i}(j)\equiv\phi_{ij}=\delta_{i,j+1}$. Using these eigenvectors the following forms for $\textbf{g}$ and $\textbf{h}$ can be obtained:
\begin{equation}
\begin{split}
g_{i,j}&=\frac{1}{2}\delta_{i,j}-\frac{1}{2}\delta_{i,j+1}+\frac{1}{2}\delta_{i,1}\delta_{j,l}\\
h_{i,j}&=\frac{1}{2}\delta_{i,j}+\frac{1}{2}\delta_{i,j+1}-\frac{1}{2}\delta_{i,1}\delta_{j,l}
\end{split}
\end{equation}
and the corresponding $\textbf{U}$ is:
\begin{eqnarray}\label{mat_U2}\
\textbf{U}= \begin{pmatrix}
\textbf{g} & \textbf{h}\\
\textbf{h} & \textbf{g}\\
\end{pmatrix}.
\end{eqnarray} 
The diagonal form of $\mathcal{K}$ is then:
\begin{equation}
\mathcal{K}=\sum_k \kappa(k)\eta_k^{\dagger}\eta_k
\end{equation}
where 
\begin{equation}
\begin{split}
\kappa(k)=\left\lbrace \begin{matrix}
+1 & \ \ \text{if} &\ \  1<k\leq l\\
0 & \ \ \text{if} &\ \  k=1
\end{matrix}\right. 
\end{split}
\end{equation}
Having the $\textbf{U}$ matrix now one can write the desired entanglement Hamiltonian in the basis of $\eta$ and then use Eq. (\ref{generic-probability}) to calculate the probability of having particular number of kinks in the ground state of the quantum spin chain. We note that here we can have $n=0,1,...,l-1$ number of kinks with the degeneracies
2${l-1}\choose{n}$.

\subsection{Generating function and moments}

In this section following the same lines of thinking as above we give formulas regarding the generating function and moments of arbitrary
quadratic observables. Compared to the results in \cite{NR2017} these formulas have simpler forms.
The generating function for an arbitrary operator $\mathcal{O}_{D}$ is defined as:
\begin{eqnarray}\label{generating function Definition}
M(z)=\text{tr}\left[ \rho_De^{z\mathcal{O}_{D}}\right].
\end{eqnarray}
The trace can be calculated explicitly in the $\delta$ representation and after some manipulations the final result is 
\begin{equation}\label{generating function 2}
M(z)=\det\frac{1}{2}(\mathbb{I}-\textbf{G})\left[{\rm \det}(\textbf{F}_{s})\right]^{\frac{1}{2}}
\det\left[\mathbb{I}+e^{z\Lambda}\tilde{\textbf{T}}\right]^{\frac{1}{2}};
\end{equation}
where $\boldsymbol{\Lambda}=\begin{pmatrix}
|\boldsymbol{\lambda}| & 0\\
0 & -|\boldsymbol{\lambda}|\\
  \end{pmatrix}$ is the matrix of the eigenvalues of the observable $\mathcal{O}_{D}$. After simple expansion we have
\begin{eqnarray}\label{generating function 3}
M(z)=\det\left[\mathbb{I}+\sum_{n=1}^{\infty}\frac{z^n}{n!}\tilde{\boldsymbol{\tau}}_n\right]^{\frac{1}{2}};\\
\end{eqnarray}
where $\tilde{\boldsymbol{\tau}}_n=\boldsymbol{\Lambda}^n\frac{\tilde{\textbf{T}}}{\mathbb{I}+\tilde{\textbf{T}}}$. To calculate the moments we need to use the following formula\cite{Withers2010}
\begin{eqnarray}\label{T22}\
\det\left[\mathbb{I}+\sum_{n=1}^{\infty}\frac{z^n}{n!}\tilde{\boldsymbol{\tau}}_n\right]^{\frac{1}{2}}=1+\sum_{k=1}^{\infty}t_{k}\frac{z^{k}}{k!},
\end{eqnarray} 
where 
\begin{eqnarray}\label{T222}\
t_{k}=\sum_{j=1}^{k}f_{j}B_{kj}(g);
\end{eqnarray} 
and $f_{j}=\frac{1}{2^j}$ and  $B_{kj}(g)$ is the partial exponential Bell polynomial 
defined as
\begin{equation}\label{Bell polynomial}
\frac{1}{j!}\Big{(}\sum_{k=1}^{\infty}\frac{g_k\epsilon^k}{k!}\Big{)}^j=\sum_{k=j}^{\infty}B_{kj}(g_1,...,g_{k-j+1})\frac{\epsilon^k}{k!},\ j=0,1,2,...,
\end{equation}
where $\epsilon$ is just a parameter. Here, we list the first few terms,
\begin{subequations}
\begin{eqnarray}\label{E with respect to g}
t_0&=&1,\\
t_1&=&f_1g_1, \\
t_2&=&f_1g_2+f_2g_1^2,\\
t_3&=&f_1g_3+f_2(3g_1g_2)+f_3g_1^3,\\
t_4&=&f_1g_4+f_2(4g_1g_3+3g_2^2)+f_3(6g_1^2g_2)+f_4g_1^4,
\end{eqnarray}
\end{subequations}
where $g=(g_1,g_2,...)$ with
\begin{eqnarray}\label{g}
g_k=\sum_{j=1}^k(-1)^{j-1}(j-1)!{\rm tr} B_{kj}(\boldsymbol{\tilde{\tau}}),
\end{eqnarray}
where $\boldsymbol{\tilde{\tau}}=(\boldsymbol{\tilde{\tau}}_1,\boldsymbol{\tilde{\tau}}_2,...)$. 
The ${\rm tr} B_{kj}(\boldsymbol{\tilde{\tau}})$ can 
be calculated first by calculating $ B_{kj}(g)$ and then symmetrization of all the terms 
$G_1G_2..G_r\to  \frac{1}{r!}\sum_r G_{\pi_1}G_{\pi_2}...G_{\pi_r}$, where the $G_i$'s are any sequence of the $\{g_k\}$ and the sum is over all permutations.
After having a symmetrized form for $ B_{kj}(g)$, we can now replace $\{g_k\}$ with $\{\boldsymbol{\tilde{\tau}}_k\}$ and derive the formulas 
for ${\rm tr} B_{kj}(\boldsymbol{\tilde{\tau}})$. Here, we list a few of the coefficients,
\begin{subequations}
\begin{eqnarray}\label{g with respect to tau}
g_1&=&{\rm tr} \boldsymbol{\tilde{\tau}}_1,\\
g_2&=&{\rm tr}\left[\boldsymbol{\tilde{\tau}}_2-\boldsymbol{\tilde{\tau}}_1^2\right],\\
g_3&=&{\rm tr}\left[\boldsymbol{\tilde{\tau}}_3-3\boldsymbol{\tilde{\tau}}_1\boldsymbol{\tilde{\tau}}_2+2\boldsymbol{\tilde{\tau}}_1^3\right],\\
g_4&=&{\rm tr}\left[\boldsymbol{\tilde{\tau}}_4-4\boldsymbol{\tilde{\tau}}_1\boldsymbol{\tilde{\tau}}_3-3\boldsymbol{\tilde{\tau}}_2^2+12\boldsymbol{\tilde{\tau}}_1^2
\boldsymbol{\tilde{\tau}}_2-6\boldsymbol{\tilde{\tau}}_1^4\right],
\end{eqnarray}
\end{subequations}
Finally we have
\begin{eqnarray}\label{T222}\
E_m=\langle\mathcal{O}_{D}^m\rangle=t_m.
\end{eqnarray} 
The first two moments can be written as follows:
\begin{eqnarray}\label{explicit two moments}\
E_1&=&\frac{1}{2}{\rm tr} \boldsymbol{\tilde{\tau}}_1,\\
E_2&=&\frac{1}{2}{\rm tr}\left[\boldsymbol{\tilde{\tau}}_2-\boldsymbol{\tilde{\tau}}_1^2\right]+\frac{1}{4}{\rm tr^2} \boldsymbol{\tilde{\tau}}_1
\end{eqnarray} 
It is also possible to get simple formulas for the fluctuations around the average:
\begin{eqnarray}\label{shifted ones}\
\tilde{E}_{m}=\langle (\mathcal{O}_{D}-\langle\mathcal{O}_{D}\rangle)^{m}\rangle.
\end{eqnarray} 
The first few terms are:
\begin{eqnarray}\label{Hl1}
\tilde{E}_1&=& 0, \nonumber\\
\tilde{E}_2&=&\frac{g_2}{2}, \nonumber\\
\tilde{E}_3&=&\frac{g_3}{2}, \nonumber\\
\tilde{E}_4-3\tilde{E}_2^2&=&\frac{g_4}{2}, \nonumber\\
\tilde{E}_5-10\tilde{E}_2\tilde{E}_3&=&\frac{g_5}{2}. \nonumber\\
\end{eqnarray}
The most general case can be written as,
\begin{equation}\label{ E tilde general}
\sum_{j=1}^m(-1)^{j-1}(j-1)!B_{mj}(\tilde{E}_1,\tilde{E}_2,...,\tilde{E}_{m-j+1})=\frac{g_m}{2}.
\end{equation}
Finally one can also write the cumulants defined as:
\begin{eqnarray}\label{ comulant}
\kappa_m=\frac{d^m}{dz^m}\log M(z)|_{z=0}
\end{eqnarray}
with respect to the $E_m$ and $\tilde{E}_m$ as follows~\cite{bell1934exponential}:
\begin{equation}\label{ comulants}
\kappa_m=\sum_{j=1}^m(-1)^{j-1}\left\lbrace \begin{matrix}
(j-1)!B_{mj}(0,\tilde{E}_2,...,\tilde{E}_{m-j+1})\\
(j-1)!B_{mj}(E_1,E_2,...,E_{m-j+1})
\end{matrix}\right. 
\end{equation}
where $m>0$ for the first branch and $m>1$ for the second one. The above formulas have relatively simpler form than the ones presented in \cite{NR2017}. However, we note that here we assume that the observable should be first diagonalized
and then the moments should be calculated. In the approach of \cite{NR2017} no diagonalization is needed.

\section{Transverse field $XY$ chain with a staggered line defect}
\label{sec:4}
In this section we would like to provide an exact determinant formula for the formation probability of the Neel sub-configuration, i.e. $\left| \downarrow\uparrow\downarrow\uparrow...\right\rangle$,
in the $XY$ chain by a direct method, that is the configuration in which the spins are in the Neel state for the sites lying in the interval $[1,n]$. In principle the formulas 
(\ref{main relations a}) and (\ref{main relations b}) are explicit examples which lead to block Toeplitz matrices. However, the direct method has this advantage that one can get
an exact formula for the correlation matrix of the line defect problem which can be useful for its own sake. In this section we first explicitly show that the FP
of the Neel configuration is the EFP for the staggered $XY$ Hamiltonian. Then we solve the line defect Hamiltonian by using the J-W transformation and find the exact correlation matrix
and EFP for the ground state.  

The Hamiltonian of the transverse field $XY$ chain is:
\begin{equation}
H_{XY}=-J\sum_{l=1}^{L} \left( \frac{1+\gamma}{4}\sigma_l^x\sigma_{l+1}^x+\frac{1-\gamma}{4}\sigma_l^y\sigma_{l+1}^y\right) -\frac{h}{2}\sum_{l=1}^L\sigma_l^z,
\label{Eq:Hamiltonian}
\end{equation}
where $\sigma_l^i$ ($i=x,y,z$) are the Pauli matrices, $J$ is the exchange parameter and $h$ is the magnetic field. The FP of the Neel state for the interval of length $n$
(which we take always even) is readily found to be
\begin{equation}
\begin{split}
&p_{\uparrow\downarrow}(n)=\\
&\left\langle g\right| \left( \frac{1-\sigma_1^z}{2}\right) \left( \frac{1+\sigma_2^z}{2}\right) \left( \frac{1-\sigma_3^z}{2}\right) ...\left( \frac{1+(-1)^n\sigma_n^z}{2}\right)\left| g\right\rangle\\
&=\left\langle g'\right| \prod_{j=1}^n\frac{1-\sigma_j^z}{2}\left| g'\right\rangle,
\end{split}
\label{Eq:staggeredProbability}
\end{equation}
where $\left| g\right\rangle$ is the ground state of the $XY$ chain, $\left| g'\right\rangle\equiv P_x^{(n)} \left| g\right\rangle$, and $P_x^{(n)}$ 
is the projection operator defined by $\prod_{j=1}^{\frac{n}{2}}\sigma_{2j}^x$, satisfying the relation $\left( P_x^n\right)^2=1$. One can easily check 
that $\left| g'\right\rangle$ is the ground state of $H'_{XY}\equiv P_x^{(n)}H_{XY}P_x^{(n)}$ with the same ground state energy as $H_{XY}$. 
Therefore, $p_{\uparrow\downarrow}(n)$ is the EFP [represented by $p(n)$ defined as the probability that all spins are down] of the ground state of $H'_{XY}$. After applying $P_x^{(n)}$, 
we find that the explicit form of $H'_{XY}$ is:
\begin{equation}
\begin{split}
H'_{XY}=& -J\sum_{l=1}^{L} \left( \frac{1+\gamma}{4}\sigma_l^x\sigma_{l+1}^x+\frac{1-\gamma}{4}f_n(l) \sigma_l^y\sigma_{l+1}^y\right)\\
& -\frac{1}{2}\sum_{l=1}^Lh_n(l)\sigma_l^z,
\end{split}
\label{Eq:modifiedXY}
\end{equation}
where $f_n(l)=-1$ and $h_n(l)=(-1)^{l+1}h$ for the case $l\leq n$, and $f_n(l)=+1$ and $h_n(l)=+h$ for the case $l>n$. 
To work with the fermionic Hamiltonian corresponding to the spin chain, we use J-W transformation defined by
\begin{equation}
\begin{split}
&c_l^{\dagger}\equiv \prod_{j<l}\sigma_j^z\sigma_l^+,\\
&c_l\equiv \prod_{j<l}\sigma_j^z\sigma_l^-,
\end{split}
\label{Eq:JW}
\end{equation}
where $\sigma_l^+\equiv\frac{1}{2}\left(\sigma^x+i\sigma^y\right)$, and $\sigma_l^-\equiv\frac{1}{2}\left(\sigma^x-i\sigma^y\right)$. The transformed Hamiltonian then becomes: 
\begin{widetext}
\begin{equation}
\begin{split}
H'_{XY}=&\frac{1}{2}\sum_{l=1}^{L-1}\left[ J_n(l) \left( c_l^{\dagger}c_{l+1}-c_lc_{l+1}^{\dagger} \right) +J\gamma_n(l) \left( c_l^{\dagger}c_{l+1}^{\dagger}-c_lc_{l+1}\right)\right]-\frac{NJ}{2} \left( c_L^{\dagger}c_1+\gamma c_L^{\dagger}c_1^{\dagger}+H.C.\right) -\sum_{l=1}^Lh_n(l)c_l^{\dagger}c_l+const.,
\end{split}
\end{equation}
\end{widetext}
where $const.=\frac{1}{2}\sum_{l=1}^Lh_n(l)$, and
\begin{equation}
\begin{split}
&J_n(l)\equiv \left\lbrace \begin{matrix}
J & \text{if} \ l>n\\
\gamma J & \text{if} \ l\leq n
\end{matrix}\right. \ , \ \gamma_n(l)\equiv \left\lbrace \begin{matrix}
\gamma & \text{if} \ l>n\\
1 & \text{if} \ l\leq n
\end{matrix}\right. .
\end{split}
\label{Eq:gamma-J}
\end{equation}
Note that this Hamiltonian is identical to the free fermionic Hamiltonian corresponding to the $XY$ model (without staggered interval) 
outside the staggered interval as expected. In the above equations $H.C.$ is the Hermitian conjugate term and $N$ is 
the eigenvalue of $\hat{N}\equiv\prod_{j\leq L}\left(2c_j^{\dagger}c_j-1\right)$ (note that $c_{L+1}=-Nc_1$). In the $\sigma^z$ basis, 
if the number of down spins is odd (or equivalently the odd number of fermionic \textit{vacancies}), then $N\equiv -1$ 
(corresponding to the periodic boundary conditions in the fermionic representation, i.e. \textit{Ramond (R) sector}), 
and in the other case (even number of down spins) $N\equiv +1$ (corresponding to the antiperiodic boundary conditions, i.e. \textit{Neveu-Schwartz (NS) sector}).\\
For the $R$-sector ($N=-1$) the fermionic Hamiltonian becomes periodic as follows (ignoring the constant term): 
\begin{widetext}
\begin{equation}
\begin{split}
H'_{XY}=&\frac{1}{2}\sum_{l=1}^L\left[ J_n(l) \left( c_l^{\dagger}c_{l+1}-c_lc_{l+1}^{\dagger} \right) +J\gamma_n(l) \left( c_l^{\dagger}c_{l+1}^{\dagger}-c_lc_{l+1}\right)\right]-\sum_{l=1}^Lh_n(l)c_l^{\dagger}c_l+\frac{1}{2}\sum_{l=1}^Lh_n(l),
\label{Eq:effectiveH1}
\end{split}
\end{equation}
\end{widetext}
whereas for the $N=+1$ case, the periodicity is destroyed. To retrieve the periodicity, 
we can use the transformation $\bar{c}_l=\exp\left[ i\frac{\pi(N+1)}{2L}l\right]c_l $, which 
results in $\bar{c}_{L+1}=\bar{c}_1$. The cost of this operation is that the allowed momenta become half-integer multiplications of $\frac{2\pi}{L}$. In the 
followings we denote the fermionic operators for both cases by $c$ and $c^{\dagger}$, keeping in mind that for the R-sector the momenta should be integer multiplications of $\frac{2\pi}{L}$, whereas for the NS-sector they should be half integers. \\

To proceed in finding the staggered spin probability, it is first useful to represent the same for the EFP for the ordinary transverse field $XY$ chain which is 
well studied in the literature~\cite{Shiroishi2001,Franchini2003,Franchini2005,Viti2019}. It can 
be found using Eq.~\ref{emptiness}. To make contact with the notation of
~\cite{Franchini2005} it is useful to define the matrix $\mathbf{S}=\frac{\mathbb{I}-\mathbf{G}^T}{2}$, where 
$S_{ij}(n)=s^{(1)}_{ij}+is^{(2)}_{ij}$, $s^{(1)}_{ij}\equiv\left\langle c_ic^{\dagger}_j\right\rangle$, and $s^{(2)}_{ij}\equiv i\left\langle c_ic_j\right\rangle$. Then the 
EFP of the $XY$ model (i.e. with no staggered interval involved) is shown to be:
\begin{equation}
p(n)|_{H_{XY}}=\left|\text{Det}(\textbf{S}(n)) \right|.
\end{equation}
Additionaly for this case, using the exact forms of the correlation functions one readily finds:
\begin{equation}
\begin{split}
\textbf{S}^{\text{free}}_{j,k}(n)&=\frac{1}{2}\delta_{ij}+\frac{1}{2}\int_0^{2\pi}\frac{\text{d}q}{2\pi}\sigma(q)e^{iq(j-k)},\\
\sigma(q)&=\frac{\cos q-h-i\gamma\sin q}{\sqrt{(\cos q-h)^2+\gamma^2\sin^2q}}, 
\end{split}
\label{Eq:sigma0}
\end{equation}
from which we see that 
\begin{equation}
G_{jk}=-\int_0^{2\pi}\frac{\text{d}q}{2\pi}\sigma(q)e^{-iq(j-k)}.
\end{equation}
The matrix $\textbf{S}(n)$ for the ordinary $XY$ model is a Toeplitz matrix, for which the EFP 
as the determinant of $\textbf{S}(n)$ can be found in the thermodynamic limit using the Fisher-Hardwig technique~\cite{Franchini2005}. Also
using the fact that $a_1=a_{-1}=\frac{J}{2}$, $b_1=\frac{\gamma}{2}$, and $a_0=-h$ in Eq.~(\ref{Hamiltonian}), one readily finds
\begin{equation}
\frac{f(q)}{\left| f(q) \right|}=-\frac{J\cos q-h+i\gamma\sin q }{\Lambda(q)}=- \sigma(q)^*
\end{equation}
which (noting that the $G$ matrix is real) is compatible with Eq.~(\ref{G matrix translational invariance}). 
For the present case (with the ground state $\left| g'\right\rangle$) we note that $p_{\uparrow\downarrow}(n)|_{H_{XY}}= P(n)|_{H'_{XY}}$, 
and therefore one should find a way to diagonalize $H'_{XY}$. To this end, we use the 
procedure of Lieb \textit{et. al.}~\cite{LIEB1961407}, according to which, after writing the Hamiltonian in the 
form $ H'_{XY}=\sum_{i,j}\left[c_i^{\dagger}A_{i,j}c_j +\frac{1}{2}\left(c_i^{\dagger}B_{i,j}c_j^{\dagger}+ h.c.\right) \right]$, 
one finds the energy spectrum by diagonalizing $\textbf{C}=(\textbf{A}-\textbf{B})(\textbf{A}+\textbf{B})$. The 
details of this calculation can be found in Appendices~(\ref{SEC:diagonalization}) and ~(\ref{SEC:solution}). The matrix $\textbf{C}$ for $H'_{XY}$ has the following form: 
\begin{equation}
\begin{split}
&C_{ii}=h^2+\frac{1}{2}J^2(1+\gamma^2),\\
&C_{i,i+1}=\left\lbrace \begin{matrix}
(-1)^i hJ & \text{if} \ \ i\leq n\\
-hJ  &  \text{if} \ \ i> n
\end{matrix}\right. \\
&C_{i,i+2}=\left\lbrace \begin{matrix}
-\frac{1}{4}J^2(1-\gamma^2) & \text{if} \ \ i\leq n\\
\frac{1}{4}J^2(1-\gamma^2) & \text{if} \ \ i> n
\end{matrix}\right.
\end{split}
\label{Eq:CMatrix}
\end{equation}
and $C_{i,j}=-hJ$ if $ i=1,j=L$ or $i=L,j=1$ and $C_{i,j}=0$ for other cases. 
Following Ref.~\cite{LIEB1961407} one can find the eigenvalues 
and the eigenvectors of $C$, the eigenvalue of which is represented by $\Lambda_k^2$, being the square of the energy spectrum of the system without the line defect
(see Appendices~\ref{SEC:diagonalization} and \ref{SEC:solution} for detailes).
Since $\textbf{A}$ ($\textbf{B}$) is symmetric (antisymmetric), $(\textbf{A}-\textbf{B})(\textbf{A}+\textbf{B})$ and $(\textbf{A}+\textbf{B})(\textbf{A}-\textbf{B})$ are symmetric and 
their eigenvalues are real and the eigenvectors $\psi$ and $\phi$ (see Eqs.~(\ref{phi and psi 2a}) and~(\ref{phi and psi 2b})) can be chosen to be orthogonal. 
To find such solutions, we use the trial function:
\begin{equation}
\begin{split}
\psi_{kj}\equiv &\frac{1}{2}\left[ 1+(-1)^j\right] \left(a_1\sin km_j+a_2\cos km_j \right)+\\
&\frac{1}{2}\left[ 1-(-1)^j\right] \left(a_3\sin km_j+a_4\cos km_j \right),
\end{split}
\label{Eq:pairfunction}
\end{equation}
where $k$ labels the eigenfunctions, $j$ is the number of the sites in the real space, $m_j\equiv \left[\frac{j+1}{2} \right] $, and $a_1,a_2,a_3$, and $a_4$ are the coefficients which 
have to be fixed using the Eq.~(\ref{phi and psi 2a}). This pairing mechanism facilitates the calculations (note that $\sin$ and $\cos$ 
are exact solutions for $m_j\gg m_0\equiv\frac{n}{2}$, and $m_j\ll m_0$). The strategy is as follows: We find two kinds of solutions: one for deep inside 
the staggered interval (DISI), and the other for deep outside the staggered interval (DOSI) with different coefficients. Then 
we should glue them by fulfilling the requirements at $j=1$ and $j=n$, i.e. where the staggered interval begins and ends respectively. This has
been done in Appendix~(\ref{SEC:solution}) in details. There we show that the solutions at DISI and DOSI are the same, up to a 
phase shift $k_s\rightarrow k_s-\pi$ where $k_s=\frac{4\pi}{L}s$ is twice the real 
momentum $q_s$ which is $q_s\equiv\frac{2\pi}{L}s$ for $N=-1$ (R sector) and $q_s\equiv\frac{2\pi}{L}(s+\frac{1}{2})$ for $N=+1$ (NS sector) and $s$ runs 
over $-\frac{L}{2},-\frac{L}{2}+1...,\frac{L}{2}-1$. After applying the boundary conditions, one reaches finally to the 
following function which diagonalizes $\textbf{C}$ (see Appendix~(\ref{SEC:solution})):
\begin{equation}
\begin{split}
&\psi_{sj}=\sqrt{\frac{2}{L}}\left\lbrace\begin{matrix}
-(-1)^{m_j}\cos q_s\left[j-n\right] & j\leq n\\
\cos q_s\left[j-n\right] & j>n
\end{matrix}\right. 
\end{split}
\end{equation}
with the eigenvalues $\Lambda_q^2=(J\cos q_s-h)^2+\gamma^2\sin^2 q_s$. The other independent solution is obtained by replacing $\cos$ by $\sin$. We consider 
the above solution for $q_s\leq 0$, and the $\sin$ solution for $q_s>0$. To continue we should find the other solution ($\phi_{sj}$) which can easily be obtained using Eq.~(\ref{phi and psi 1b}), for $\Lambda_q\neq 0$:
\begin{widetext}
\begin{equation}
\begin{split}
\phi_{XY}(j\leq n)=&-\sqrt{\frac{2}{L}}\Lambda_{q_s}^{-1}\left[(-)^{m_j+a_j}h\cos q_s(j-n)\right. \\
&\left. +(-)^{m_{j-1}}\left(\frac{-1+\gamma}{2}\right)\cos q_s(j-n-1)+(-)^{m_{j+1}}\left(\frac{1+\gamma}{2}\right) \cos q_s(j-n+1) \right], \\
\phi_{XY}(j= n+1)=&-\sqrt{\frac{2}{L}}\Lambda_{q_s}^{-1}\left[h\cos q_s(j-n)\right. \\
&\left. (-)^{m_{j-1}}\left( \frac{-1+\gamma}{2}\right) \cos q_s(j-n-1)-\frac{1+\gamma}{2}\cos q_s(j-n+1) \right], \\
\phi_{XY}(j> n+1)=&-\sqrt{\frac{2}{L}}\Lambda_{q_s}^{-1}\left[h\cos q_s(j-n)\right. \\
&\left. +\frac{-1+\gamma}{2}\cos q_s(j-n-1)-\frac{1+\gamma}{2}\cos q_s(j-n+1) \right], 
\end{split}
\end{equation}
\end{widetext}
where $a_j$ (not to be confused with the coefficients $a_1$, $a_2$, $a_3$ and $a_4$) is $1$ if $j$ belongs to the first sublattice (odd $j$s) and zero for the other sublattice (even $j$s). This solution is reserved for $s\leq 0$, and for $q_s> 0$ one should replace $\cos$ by $\sin$. For $\Lambda_k=0$ the solution is $\phi(j)=\pm \psi(j)$.
Having $\psi$ and $\phi$ solutions in hand, one can directly calculate $g_{si}\equiv \frac{1}{2}\left( \psi_{si}+\phi_{si}\right) $ and $h_{si}\equiv \frac{1}{2}\left(\psi_{si}-\phi_{si}\right)$ to diagonalize $H$ [see relation~\ref{Hdiag}]. For example, $\left\langle c_ic_j\right\rangle=\sum_sg_{si}h_{sj}$ where $s$ is the integer (half integer) for $N=-1$ $(N=+1)$. Working out with $f_{AB}^{i,j}=\sum_{s=-\frac{L}{2}}^{L/2-1}A_{si}B_{sj}$, in which $A,B=\psi, \phi$, one can easily show that always (irrespective to the amount of $i$ and $j$ being inside or outside the staggered interval) $f_{\psi\psi}^{i,j}=f_{\phi\phi}^{i,j}=\delta_{ij}$, and $\delta_{ij}$ is the Kronecker delta. These functions help us to calculate the important correlation functions:\\
\begin{equation}
\begin{split}
&\left\langle c_ic_j^{\dagger} \right\rangle =\frac{1}{4}\left[ f_{\psi\psi}+f_{\psi\phi}+f_{\phi\psi}+f_{\phi\phi}\right],\\ 
&\left\langle c_ic_j \right\rangle =\frac{1}{4}\left[f_{\psi\psi}-f_{\psi\phi}+f_{\phi\psi}-f_{\phi\phi}\right], \\
&\left\langle c_i^{\dagger}c_j^{\dagger} \right\rangle=\frac{1}{4}\left[f_{\psi\psi}+f_{\psi\phi}-f_{\phi\psi}-f_{\phi\phi}\right], \\ 
&\left\langle c_i^{\dagger}c_j \right\rangle =\frac{1}{4}\left[f_{\psi\psi}-f_{\psi\phi}-f_{\phi\psi}+f_{\phi\phi}\right],
\end{split}
\label{Eq:correlations0}
\end{equation}

We now calculate the correlation functions explicitly. In what follows, 
we consider the case in which $i,j\leq n$, and extension to the other cases is straightforward. To calculate
the correlation functions, we need the following identity that has been proved in Appendix~(\ref{SEC:corr_func}): 
\begin{equation}
\begin{split}
\frac{1}{4}\left( f_{\psi\phi}\pm f_{\phi\psi}\right) =\chi_{ij}^{\pm}\sigma_1(i,j)+\chi_{ij}^{\mp}\sigma_2(i,j),
\end{split}
\end{equation}
where $\chi_{ij}^{+}=\left[\frac{1+(-)^{a_i-a_j}}{2} \right](-)^{m_{j+1}-m_i}$ and $\chi_{ij}^-=\left[\frac{1-(-)^{a_i-a_j}}{2} \right](-)^{m_{j+1}-m_i}$, and also 
\begin{equation}
\begin{split}
\sigma_1(j,k)&=\frac{1}{2L}\sum_s\cos q_s(k-j)\left( \frac{-h+\cos q_s}{\Lambda_s}\right)\\
&=\frac{1}{2L}\sum_s\left( \frac{-h+\cos q_s}{\Lambda_s}\right)e^{-iq_s(k-j)},\\
\sigma_2(j,k)&=\frac{1}{2L}\sum_s\sin q_s(k-j)\left( \frac{\gamma\sin q_s}{\Lambda_s}\right)\\
&=\frac{i}{2L}\sum_s\left( \frac{\gamma\sin q_s}{\Lambda_s}\right)e^{-iq_s(k-j)},
\end{split}
\label{Eq:sigma}
\end{equation} 
where we have used the symmetry considerations to add extra zero contributions, and for 
the summation $s$ is integer for the R-sector, and half integer for the NS-sector. Therefore, if $i$ and $j$ belong to the same sublattice then we have:
\begin{equation}
\begin{split}
&\Sigma_1(i,j)\equiv\frac{1}{4}\left( f_{\psi\phi}+ f_{\phi\psi}\right)=(-)^{m_{j+1}-m_i}\sigma_1(j-i), \\
&\Sigma_2(i,j)\equiv\frac{1}{4}\left( f_{\psi\phi}- f_{\phi\psi}\right)=(-)^{m_{j+1}-m_i}\sigma_2(j-i),
\end{split}
\end{equation}
and
\begin{equation}
\begin{split}
&\Sigma_1(i,j)=(-)^{m_{j+1}-m_i}\sigma_2(j-i), \\
&\Sigma_2(i,j)=(-)^{m_{j+1}-m_i}\sigma_1(j-i),
\end{split}
\end{equation}
if they belong to the different sublattices. Therefore, at this stage we can find the explicit form of the correlation functions, which are
\begin{equation}
\begin{split}
\left\langle c_ic_j^{\dagger} \right\rangle &=\frac{1}{2}\delta_{ij}+\chi_{ij}^+\sigma_1(i,j)+\chi_{ij}^-\sigma_2(i,j),\\
\left\langle c_ic_j \right\rangle &=\chi_{ij}^+\sigma_2(i,j)+\chi_{ij}^-\sigma_1(i,j).
\end{split}
\end{equation}
Also note that $\left\langle c_i^{\dagger}c_j^{\dagger} \right\rangle =
-\left\langle c_ic_j \right\rangle$ and $\left\langle c_i^{\dagger}c_j \right\rangle =
\delta_{ij}-\left\langle c_ic_j^{\dagger} \right\rangle$, that can be readily checked. 

Finally we turn to the calculation of the formation probability of the Neel configuration which
is the EFP for $H'_{XY}$. It can be determined by calculating $|\text{Det}(\textbf{S}_n)|$, the elements of which are
$s_{ij}=\left\langle c_ic_j^{\dagger} \right\rangle - \left\langle c_ic_j \right\rangle $ as outlined
above. Using the above correlation functions, one simply obtains:
\begin{equation}
s_{jk}=\frac{1}{2}\delta_{jk}+\frac{1}{2}(\chi_{jk}^+-\chi_{jk}^-)\sigma(j,k),
\label{Eq:sMatrix}
\end{equation}
where
\begin{equation}
\begin{split}
\sigma(j,k)&\equiv 2\left( \sigma_1(j,k)-\sigma_2(j,k)\right) \\
&=\frac{1}{L}\sum_{s=-L/2}^{L/2-1}\left( \frac{\cos q_s-h-i\gamma\sin q_s}{\Lambda_s}\right)e^{iq_s(j-k)}.
\end{split}
\label{Eq:sigma2}
\end{equation}
This matrix, in the $L\rightarrow \infty$ limit becomes
\begin{equation}
\begin{split}
&\sigma(j,k)=\int \frac{\text{d}q}{2\pi} \sigma(q)e^{iq(j-k)},
\end{split}
\end{equation}
where $\sigma(q)$ is defined in Eq.~(\ref{Eq:sigma0}). $s_{jk}$ is compatible with the result for the free case, i.e. Eq.~(\ref{Eq:sigma0}), except that here
a sign matrix ($\text{sgn}(j,k)\equiv\chi_{jk}^+-\chi_{jk}^-$) is multiplied. The closed form of this sign matrix is:
\begin{equation}
\begin{split}
\text{sgn}(j,k)=&\cos\pi(k-j)\left\lbrace -(-)^{\frac{j+k}{2}}\left| \cos \frac{\pi}{2}(k-j)\right|\right.  \\
&\left. +(-)^{\frac{j-k-1}{2}}\left| \sin \frac{\pi}{2}(k-j)\right| \right\rbrace,
\end{split}
\label{Eq:sgnMatrix}
\end{equation}
For example, the explicit form for the above sign matrix for $n=8$ is

\begin{equation}
\begin{split}
\textbf{sgn} =\left[ \begin{matrix}
+ & - & - & + & + & - & - & + \\
+ & - & - & + & + & - & - & + \\
- & + & + & - & - & + & + & -\\
- & + & + & - & - & + & + & -\\
+ & - & - & + & + & - & - & + \\
+ & - & - & + & + & - & - & + \\
- & + & + & - & - & + & + & -\\
- & + & + & - & - & + & + & -
\end{matrix} \right].
\end{split}
\end{equation}
One can easily check that the above sign matrix is different from the ones suggested in (\ref{sing matrices}). As we already discussed in Sec. II
there are different sign matrices that lead to the same FP's but they come from different correlation matrices. It is worth mentioning that
since the Hamiltonian of the transverse field $XY$ chain that we considered here was with PBC we ended up having R and NS sectors for the fermionic counterparts. Finding
which one is the actual ground state of the spin system is a nontrivial problem. Since our line defect problem has the same 
eigenvalues the problem is similar to the clean case and we refer to \cite{DePasquale2009}  for systematic study of the clean case.
Note that the calculated sign matrix is correct not only for the ground state of R and NS sectors of the $XY$ chain but also for the generic one dimensional translational invariant free fermions.\\

Here we summarize the results of this section containing the final expressions. The aim of this section was to obtain the staggered (Neel) configuration probability for the ground state of the $XY$ model:
\begin{equation}\label{Eq:staggeredMain1}\
\begin{split}
p_{\uparrow\downarrow}(n)=
|\text{Det}(\textbf{S}_n)|
\end{split}
\end{equation}
where the elements of $\textbf{S}_n$ are $s_{jk}$. Using the method outlined in~\cite{LIEB1961407} to find the energy eigenvalues, we found the analytic expression for the elements of the matrix $\textbf{S}_n$, i.e.
\begin{equation}\label{Eq:staggeredMain2}
\begin{split}
s_{jk}=\frac{1}{2}\delta_{jk}+\frac{1}{2}\text{sgn}(j,k)\sigma(j,k)
\end{split}
\end{equation}
where
\begin{equation}\label{Eq:staggeredMain3}
\begin{split}
&\text{sgn}(j,k)=\cos\pi(k-j)\left\lbrace -(-)^{\frac{j+k}{2}}\left| \cos \frac{\pi}{2}(k-j)\right|\right. \\
&\left.  +(-)^{\frac{j-k-1}{2}}\left| \sin \frac{\pi}{2}(k-j)\right| \right\rbrace,
\end{split}
\end{equation}
and
\begin{equation}\label{Eq:staggeredMain4}
\sigma(j,k)=\frac{1}{L}\sum_{s=-L/2}^{L/2-1}\left( \frac{\cos q_s-h-i\gamma\sin q_s}{\Lambda_s}\right)e^{iq_s(j-k)}.
\end{equation}
As we mentioned before, Eqs. (\ref{Eq:staggeredMain1}), (\ref{Eq:staggeredMain2}) and (\ref{Eq:staggeredMain3}) are valid for all the one-dimensional translationally invariant free fermions as far as we use the corresponding $\sigma(j,k)$.


\section{Probability distribution of particle numbers and kinks in the transverse field $XY$ chain}
\label{sec:5}
In this section we provide a couple of examples to show how the explicit formulas that we provided in the previous sections can be applied to
calculate the probability distribution of quadratic observables. In both cases the model that we take is the ground state of the transverse field $XY$ chain which has a rich phase diagram with 
three critical lines at $h=1$, $\gamma\neq0$; $h=-1$, $\gamma\neq0$ and $\gamma=0$, $-1<h<1$. Here we concentrate mostly on the non-negative transverse field
part of the phase diagram and study probability distribution of particle numbers and kinks in the transverse field $XY$ chain.
\subsection{Probability distribution of magnetization}
\label{sec:5A}
\begin{figure}[t]
	\centering
	\includegraphics[width=0.55\textwidth]{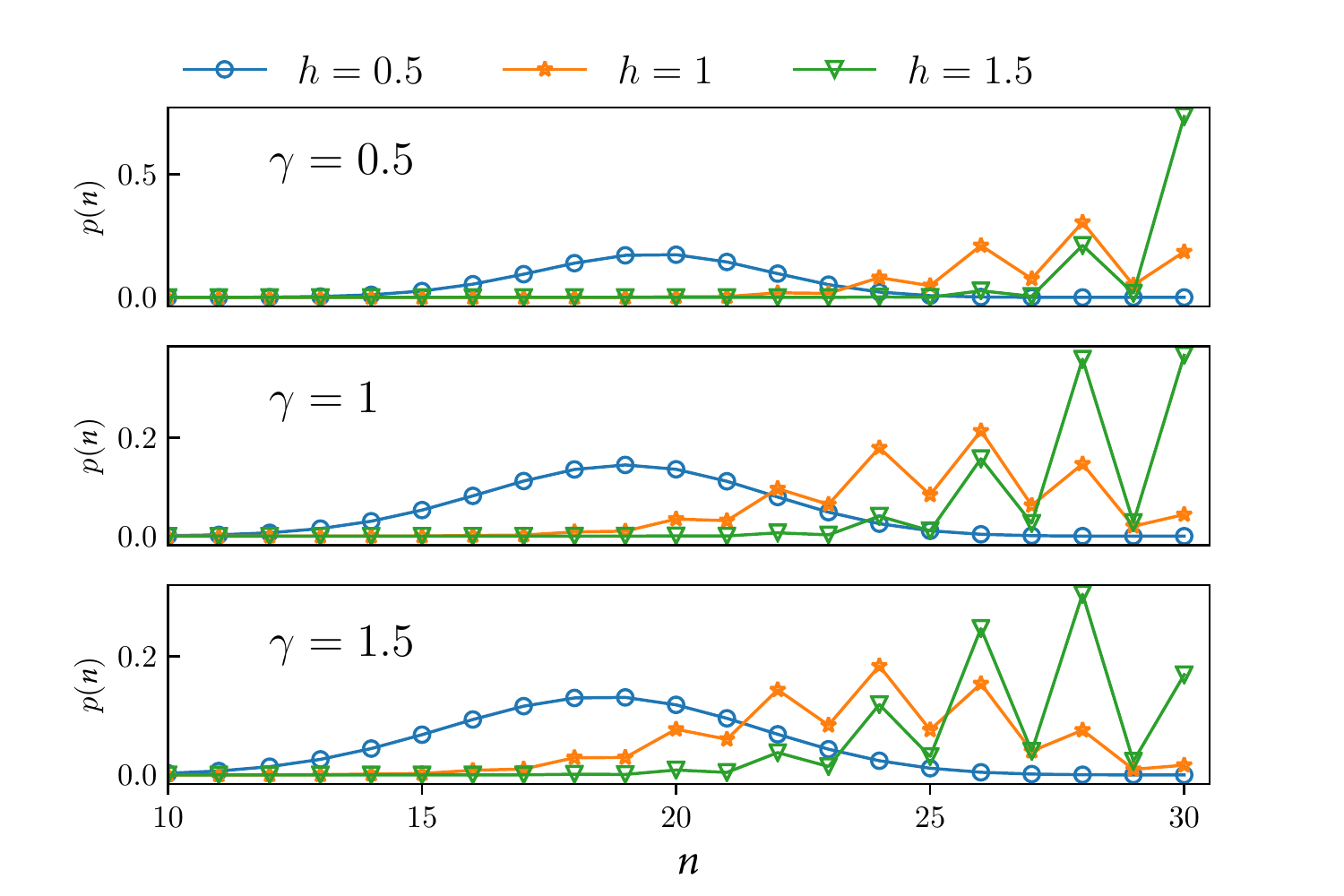}
	\caption{\footnotesize Probability distribution of the number of particles, equivalent to the magnetization in the $\sigma^z$ direction in terms of $\gamma$ and $h$ for the $XY$-chain with $L=200$ and $l=30$.}
	\label{fig:1}
\end{figure}

\begin{figure}[t]
	\centering
	\includegraphics[width=0.55\textwidth]{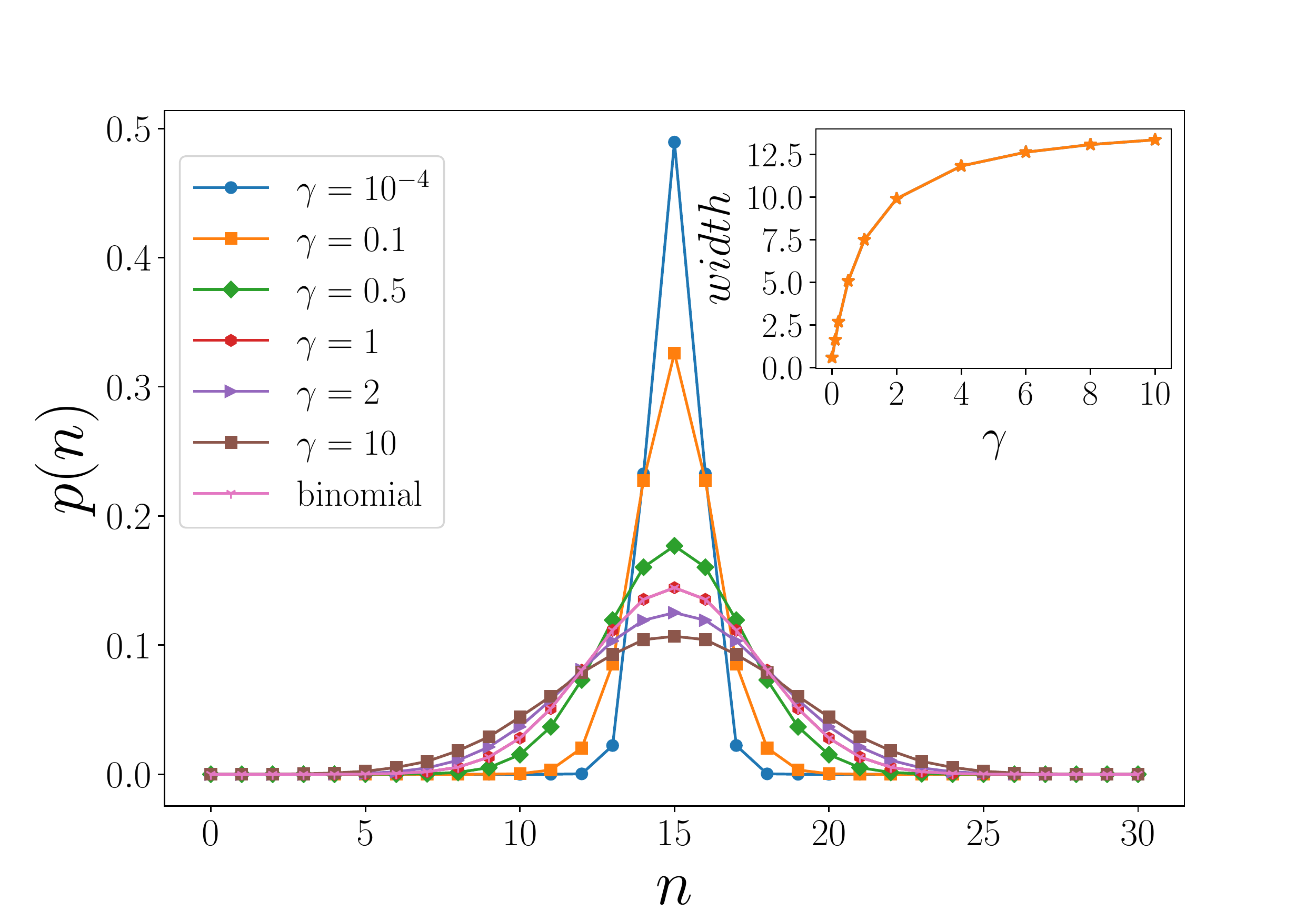}
	\caption{\footnotesize Probability distribution of the number of particles, equivalent to the magnetization in the $\sigma^z$ direction for the $XY$ chain with $L=200$ and $l=30$, in terms of $\gamma$ for $h=0$. The inset shows the width of the distribution ($\left\langle \mathcal{N}^2 \right\rangle-\left\langle \mathcal{N}\right\rangle^2$), obtained from the Eq~\ref{Eq:width}.}
	\label{fig:4}
\end{figure}

The first example is the distribution of the magnetization in the $\sigma^z$ direction in an interval of size $l$. This is equivalent to the distribution
of the number of particles that we have studied in the Sec.~\ref{sec:stat-number} by the relation $\mathcal{M}_z\equiv\sum_i \sigma^z_i=2\mathcal{N}-1$, where $\mathcal{N}$ is the total number operator of Fermions in the subsystem, with eigenvalue $n$. The generating function of this quantity in the thermodynamic limit
has been already studied in \cite{Ivanov2013}, however, it does not seem to be straightforward to do the inverse Laplace transform analytically or numerically in
the most generic cases. Using the formulas of the Sec.~\ref{SEC:kinks} with appropriate $\textbf{G}$ and $\textbf{F}$
matrices we can easily calculate this distribution numerically for arbitrary parameters of the Hamiltonian.
The results are depicted in Fig.~\ref{fig:1} which shows clear change of behavior when we cross the critical line $h=1$. The emergent oscillations in the region $h\geq1$ are similar to the ones that have already been seen in the study of the EFP in \cite{Franchini2005} and attributed to the competition between the energy cost of flipping a spin (controlled by $h$) and the superconducting terms which create and destroy fermions in pairs (controlled by $\gamma$). Although not shown here, similar 
oscillations also appear in the region $-1\geq h$ with peaks shifted to the left part of the graph. \\
To better understand these results we first consider $h=0$ (other $h<1$ values are shifts of the graphs to right, keeping the shape of the graph nearly unchanged). For $\gamma=1$ we have the Ising model with $Z_2$ symmetry, for which the ground state is
\begin{equation}
\left| g\right\rangle =\frac{1}{\sqrt{2}}\left[\left|+,+,...,+ \right\rangle_x+\left|-,-,...,- \right\rangle_x  \right]
\end{equation}
where $\left|\pm\right\rangle_x$ is the eigenstate of $\sigma^x$ with $\pm1$ eigenvalue. Using this, one can easily find the reduced density matrix:
\begin{equation}
\begin{split}
\rho_l =\frac{1}{2^{l+1}}&\sum_{\left\lbrace \sigma_z\right\rbrace \left\lbrace \sigma'_z\right\rbrace }\left(1+(-)^{n_-^{\sigma}+n_-^{\sigma'}} \right) \times\\
&\left|\sigma_z^{(1)},\sigma_z^{(2)}...\sigma_z^{(l)}\right\rangle \left\langle {\sigma'_z}^{(1)},{\sigma'_z}^{(2)}...{\sigma'_z}^{(l)} \right| 
\end{split}
\end{equation}
where $n_-^{\sigma}$ is the number of down spins, and the summations are over all spin configurations $\sigma_z$ and $\sigma'_z$. The formation probabilities, which are the diagonal elements of the above equation are therefore constant, i.e. $\frac{1}{2^l}$. Therefore we see that \textit{all} configurations appear in the reduced density matrix with the same probability, i. e. for the zero-magnetic field Ising model the probability distribution of particle numbers is expected to be bimonial, which is 
\begin{equation}
p_{\gamma=1,h=0}(n)=p_{\text{binomial}}=\frac{1}{2^l}\frac{l!}{(l-n)!n!}
\end{equation}
In Fig.~\ref{fig:4} we show the results for the $XY$ model with magnetic field fixed to zero. All distributions have a mean value at $\frac{l}{2}$. We see that the graph for $\gamma=1$ fits completely to the binomial distribution as expected. The amount of $\gamma$ controls the width of the distribution, so that for $\gamma=0$ we have a $U(1)$ symmetry, and the number of particles is fixed and consequently
the distribution is just a Dirac delta function. For small $h$ the effect of increasing $\gamma$ [which controls the strength of $U(1)$ symmetry breaking] is just broadening the distribution by increasing the variance. It is easy to show that (see, for example, \cite{NR2017}) the average of the number of fermions is given by :
\begin{equation}
\left\langle \mathcal{N}\right\rangle=\frac{l}{2}\left(1 +l^{-1}\text{Tr}\left[G \right]\right) 
\label{Eq:average}
\end{equation}
and its fluctuation is ($\mathcal{N}'\equiv l^{-\frac{1}{2}}\mathcal{N}$):
\begin{equation}
\left\langle \mathcal{N}'^2 \right\rangle-\left\langle \mathcal{N}'\right\rangle^2=\frac{1}{4}\left(1 -l^{-1}\text{Tr}\left[G^2 \right]\right) 
\label{Eq:width}
\end{equation}
where the matrix $G$ was defined in Eq.~(\ref{Eq:G}) (note that for $\gamma=1$ and $h=0$ we have $G=0$, so that
$\left\langle \mathcal{N}'^2 \right\rangle-\left\langle \mathcal{N}'\right\rangle^2=\frac{1}{4}$). The variance of $\mathcal{N}$ is shown in the inset of Fig.~\ref{fig:4} for $h=0$, in which we see that the width of the distribution function increases with $\gamma$. Generally, in the thermodynamic limit, this function tends to a value that depends on the $\gamma$ and $h$.

\begin{figure}[t]
	\centering
	\includegraphics[width=0.55\textwidth]{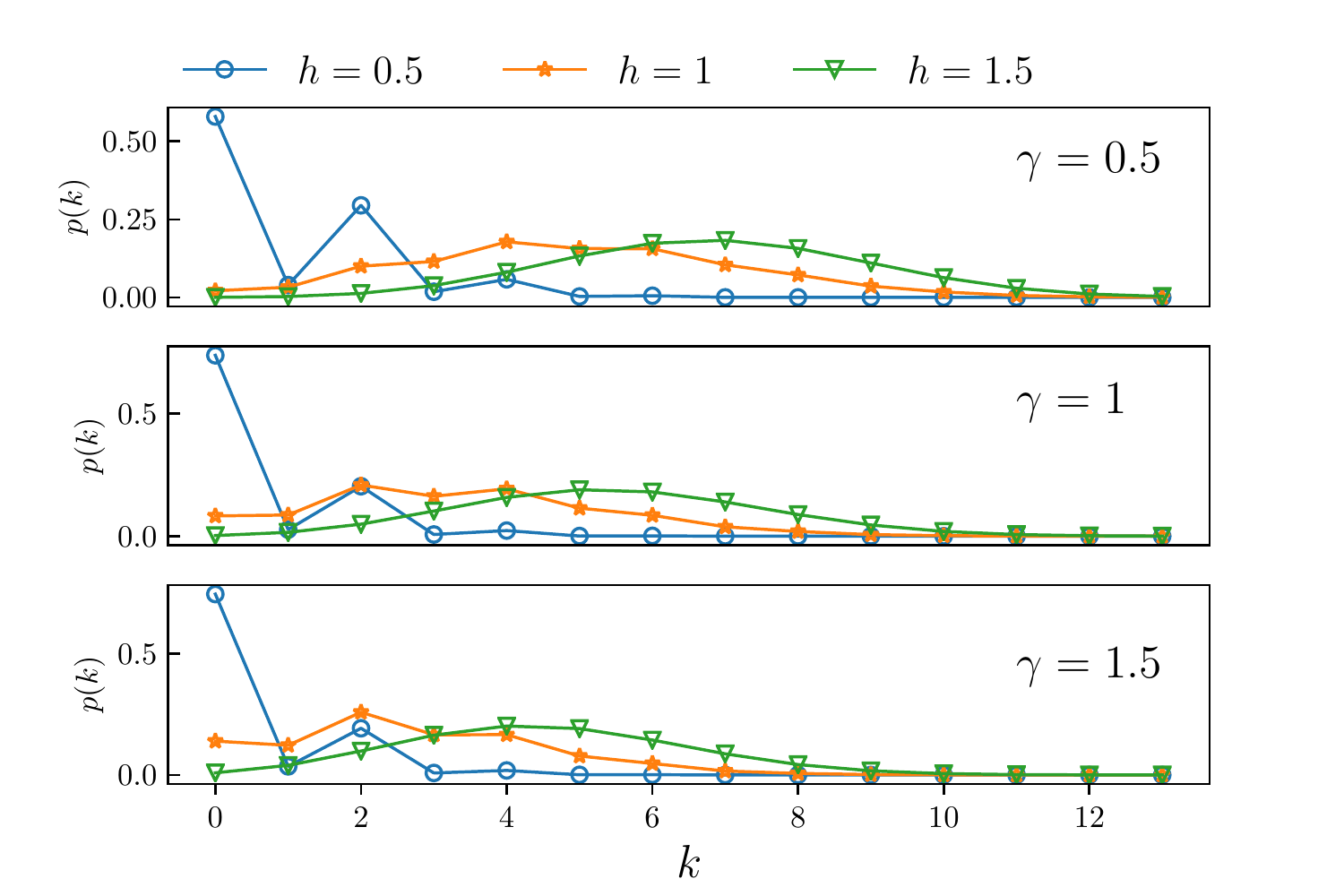}
	\caption{\footnotesize Probability distribution of the kinks in the $\sigma^x$ direction for the $XY$ chain ($L=200$ and $l=18$) using the exact diagonalization outlined in SEC~\ref{sec:3}. The graph shows the dependence on $\gamma$ and $h$.}
	\label{fig:2}
\end{figure}

\begin{figure}[t]
	\centering
	\includegraphics[width=0.50\textwidth]{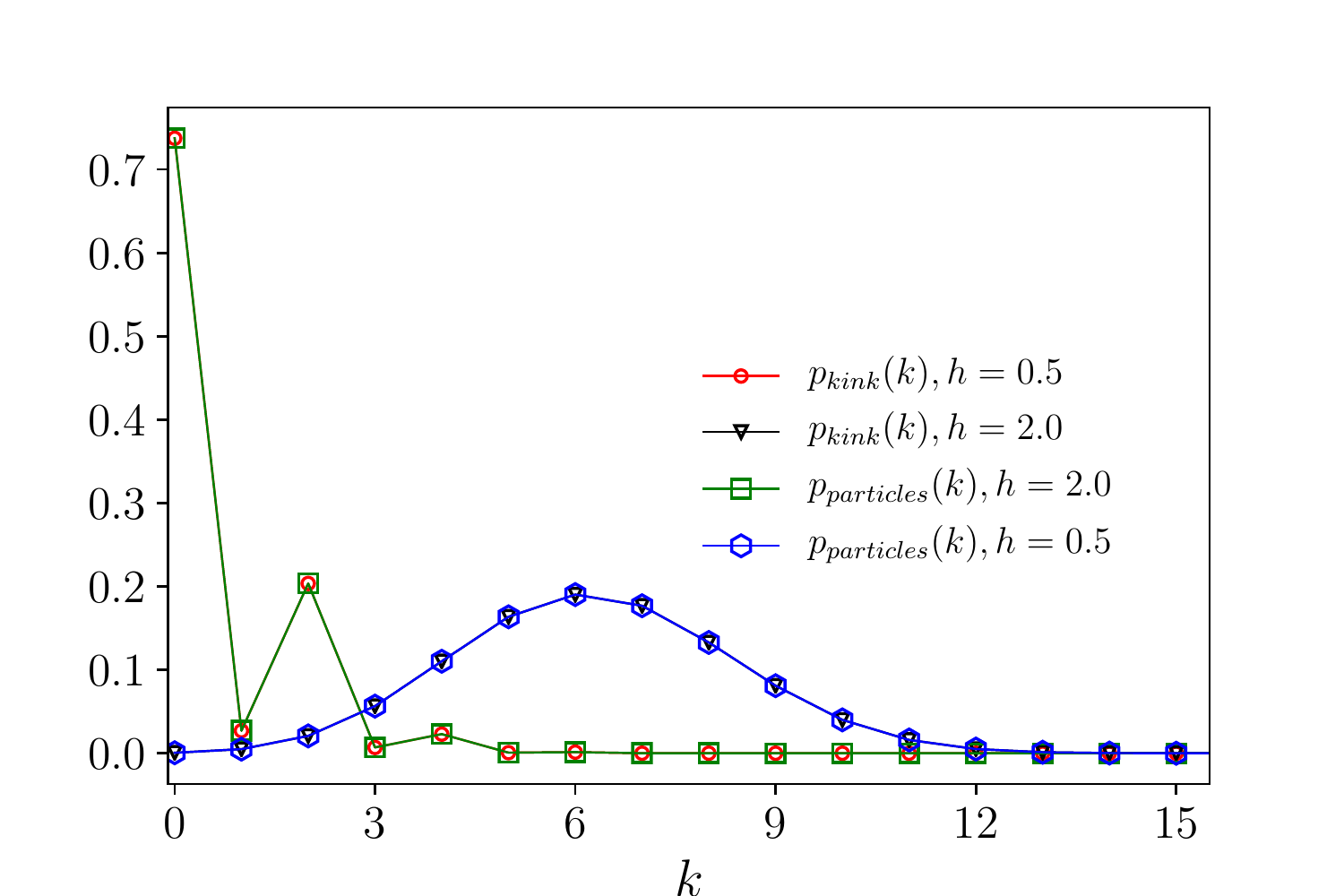}
	\caption{\footnotesize Illustration of the Kramers-Wannier duality. The figure shows the probability distribution of the magnetization ($p_{particles}$) and the kinks ($p_{kink}$) for $\gamma=1$ and $h=2$ and $\tilde{h}=\frac{1}{h}=0.5$ for $L=200$ and $l=18$.}
	\label{fig:3}
	\end{figure}

\subsection{Probability distribution of kinks}
\label{sec:5B}
 As a second example of our formalism we discuss the distribution of kinks in the ground state of the $XY$ chain. Using 
 the method that was provided in Sec.~\ref{SEC:kinks}, we calculated with exact numerical calculations 
 the probability of having different number of kinks in the ground state and presented the results in 
 the Figure~\ref{fig:2}. Similar to the particle number distribution here too we have clear change 
 of behavior around the critical line. However, the oscillations are now appearing in the regions $-1\geq h\geq 1$. This 
 is not surprising because the $\sigma^z_j$ and $\sigma^x_j\sigma^x_{j+1}$ have dual behavior. In the case of the 
 transverse field Ising chain this duality is exact and it is called Kramers-Wannier(KW) duality which connects the 
 Hamiltonian with the magnetic field  $h$  to the one with the magnetic field $\tilde{h}\equiv\frac{1}{h}$. To see the 
 effect of this duality on the distribution we plotted in Fig.~\ref{fig:3} the particle number and kink distribution 
 for two different dual magnetic fields. To make the comparison easy we mirrored and shifted (by one unit) one of the 
 distributions. The two distributions are perfectly matching which is a nice way to see the effect of the KW duality on the FCS. \\
 One can get some intuition about the fluctuations of the kink probability distribution using a similar argument as the previous section. Using the same equations~\cite{NR2017} for the kink operator (Eq.~\ref{Eq:kink}) one easily finds that:
 \begin{equation}
 \left\langle \mathcal{K}\right\rangle=\frac{l-1}{2}\left( 1+\frac{1}{l-1}\sum_{j}G_{j,j+1}\right)
 \end{equation}
 and also ($\mathcal{K}'\equiv (l-1)^{-\frac{1}{2}}\mathcal{K}$)
 \begin{equation}
 \left\langle \mathcal{K}'^2 \right\rangle-\left\langle \mathcal{K}'\right\rangle^2=\frac{1}{4}\left( 1-\frac{1}{l-1}\sum_{jk}G_{j,k+1}G_{k,j+1}\right) 
 \label{Eq:width-kink}
 \end{equation} 
The second term in the large $l$ limit gives a number which depends on the parameters of the Hamiltonian. This number can be in principle written as a double integral. When $\gamma=0$ and $h>1$, $p(k)=\frac{1}{2^{l-1}}$${l-1}\choose{k}$ 
is completely symmetric, having its peak at $k_{\text{max}}=\frac{l-1}{2}$. As $\gamma$ increases, this peak shifts to the left, i.e. lower amounts. But this does not continue to zero kink, i.e. for large $\gamma$s both the average and the width of kinks saturates. To see the properties of kinks at large $\gamma$s let us inspect the $XY$ Hamiltonian~\ref{Eq:Hamiltonian}, which becomes $H_{XY}(\gamma\rightarrow\infty)\approx -\frac{\gamma J}{4}\sum_{l}\left( \sigma_l^x\sigma_{l+1}^x-\sigma_l^y\sigma_{l+1}^y\right)$, including two competing terms. The first term (which commutes with $\mathcal{K}$) is minimal in large $\gamma$s when there is no kink, whereas the second term generates fluctuations preventing the ground state from having a number of kinks less than a minimal value. To see this, let us consider only the second term the ground state of which is
\begin{equation}
\left| g\right\rangle_y=\frac{1}{\sqrt{2}}\left[\left|+,-,+,-... \right\rangle_y+\left|-,+,-,+... \right\rangle_y \right] 
\end{equation}
where $\left|\pm \right\rangle_y$ is the eigenvector of $\sigma^y$ with the eigenvalue $\pm 1$. By constructing the reduced density matrix in the basis of $\sigma^x$, one can show that the probability of having $k$ kinks is
\begin{equation}
 p_k=\frac{1}{2^{l-1}} {{l-1}\choose{k}}+\frac{1}{2^{l-1}}\sum_{j=0}^{\frac{l}{2}}(-1)^{k-j}{{\frac{l}{2}-1}\choose{j}}{{\frac{l}{2}}\choose{k-j}}
 \label{Eq:kink-distributionY}
 \end{equation} 
Using this relation, we find that $\left\langle \mathcal{K}^2\right\rangle=\frac{l(l-1)}{4}$ and $\left\langle \mathcal{K}\right\rangle=\frac{l-1}{2}$, which gives the width of the distribution $\frac{l-1}{4}$. Therefore we see that the second term generates kinks, so that we expect that in the limit $\gamma\rightarrow\infty$ the width of the distribution of kinks becomes a finite value. Another way to understand this is to look at the kink operator, which commutes with $\sum_j\sigma_j^x\sigma_{j+1}^x$, so that they have simultaneous eigenvectors, e.g. a state with zero kinks which is the ground state of the first term of $H_{XY}(\gamma\rightarrow\infty)$. The second term of $H_{XY}(\gamma\rightarrow\infty)$ (containing $\sum_j\sigma_j^y\sigma_{j+1}^y$) does not commute with $\mathcal{K}$, which according to the Heisenberg uncertainty relation generates uncertainty in the expectation value of $\mathcal{K}$. This translates to generating a finite width in the distribution of kinks in the $\gamma\rightarrow\infty$ limit.


\section{Conclusions}
In this paper we studied formation probabilities and full counting statistics of the quadratic observables in the free fermions and the corresponding spin chains.
We first showed that the problem of FP of the ground state of a generic free fermion
can be translated into an emptiness formation probability of a free fermionic Hamiltonian with defects. In one dimension the defect is a line but in higher dimensions it can have different
forms. Using the same line of thinking we then provided determinant formulas for the FP's with respect to the correlation matrix of the ground state of
the Hamiltonian. In the second part of the paper we studied FCS of a generic quadratic observable in the ground state of a generic free fermion Hamiltonian. We showed that the probability
of finding a particular value for the observable is exactly a FP problem for a Hamiltonian written in the basis that diagonalizes the observable. We showed how 
this can be done for a full system and also for the subsystem in the most generic case. Two simple cases, i.e. fluctuations of particles and kinks were discussed to show how one should implement
the presented ideas.
Finally, in the last section we solved the problem of the transverse field $XY$ chain with a staggered magnetic line defect. We found exact correlation functions in and outside of the 
staggered region and provided a determinant formula for the FP of the staggered configuration in the ground state of the $XY$ chain. Throughout the paper we tried to keep the discussion  
general except when we were presenting explicit examples to show how the procedure can be followed. Clearly one can take a particular model such as the $XY$ chain and apply the presented
methods. The numerical method to calculate these quantities are quite simple, however, to further push the analytical calculations in specific cases one normally needs to hire such methods as 
the generalized Fisher-Hartwig theorem which is beyond the scope of this paper and we hope to come back to them in future works. 

Finally we stress that all of our results are valid independent of dimension as far as we have Wick's theorem and 
$\delta_{jk}=\langle \gamma_j\gamma_k\rangle=-\langle \bar{\gamma}_j\bar{\gamma_k}\rangle$ and $iG_{jk}=\langle \bar{\gamma}_j\gamma_k\rangle$ which is the case for all the eigenstates of quadratic Hamiltonians with real couplings. The same can be also applied for the finite temperature systems and for the generalized Gibbs ensemble.
When Wick's theorem is valid but $\langle \gamma_j\gamma_k\rangle$ and $\langle \bar{\gamma}_j\bar{\gamma_k}\rangle$ are not identity matrices like when the couplings are not real or for the time-dependent cases, one needs to make a small adjustment to the above equations that we leave for future studies.
\newline
\newline

\textbf{Acknowledgements.}
MAR acknowledges support from CNPQ and FAPERJ (grant number 210.354/2018). We thank P. Calabrese for early discussions.

\appendix
\section{Diagonalization of the Free Fermions}\label{SEC:diagonalization}
\label{app1}
In this subsection, we summarize the result of Ref.\cite{LIEB1961407}. Consider a generic quadratic observable with real couplings
\begin{equation}\label{Hsub}\
\textbf{O}=\sum_{ij}\left[c_{i}^{\dagger}M_{ij}c_{j}+\frac{1}{2}c_{i}^{\dagger}N_{ij}c_{j}^{\dagger}+\frac{1}{2}c_{i}N_{ji}c_{j}\right]-\frac{1}{2}{\rm Tr}{\textbf{M}},
\end{equation} 
%
where $c_{i}^{\dagger}$ and $c_{i}$ are fermionic creation and annihilation operators, and $i$ and $j$ run over the interval $\left\lbrace 1,2,...,L\right\rbrace $.  The Hermitian observable requires $\textbf{M}$ and $\textbf{N}$ to be symmetric and antisymmetric matrices respectively. To diagonalize the operator we use the following canonical transformation
\begin{eqnarray}\label{mat_U1}\
\begin{pmatrix}
\textbf{c} \\
\textbf{c}^{\dagger} \\
\end{pmatrix}=
\textbf{U}^{\dagger}
\begin{pmatrix}
\boldsymbol{\eta} \\
\boldsymbol{\eta}^{\dagger} \\
\end{pmatrix},
\end{eqnarray} 
with 
\begin{eqnarray}\label{mat_U2}\
\textbf{U}= \begin{pmatrix}
\textbf{g} & \textbf{h}\\
\textbf{h}^{*} & \textbf{g}^{*}\\
\end{pmatrix}.
\end{eqnarray} 
which results in $g$ and $h$ being $L\times L$ matrices, and the diagonal form of $\textbf{O}$ is
\begin{eqnarray}\label{Hdiag}\
\textbf{O}=\sum_{k}|\lambda_{k}|(\eta_{k}^{\dagger}\eta_{k}-\frac{1}{2}).
\end{eqnarray} 
By requiring that $[\eta_k,\textbf{O}]=|\lambda_{k}|\eta_k$, it is found that:
\begin{equation}
\begin{split}
& \eta_kg_{ki}=\sum_j \left(g_{kj}M_{ji}-h_{kj}N_{ji} \right)\\
& \eta_kh_{ki}=\sum_j \left(g_{kj}N_{ji}-h_{kj}M_{ji} \right)
\end{split}
\label{Eq:g and h}
\end{equation}
By defining new matrices $\boldsymbol{\psi}$ and $\boldsymbol{\phi}$ as follows
\begin{eqnarray}\label{gandh}\
\textbf{g}&=&\frac{1}{2}(\boldsymbol{\psi}+\boldsymbol{\phi}),\\
\textbf{h}&=&\frac{1}{2}(\boldsymbol{\psi}-\boldsymbol{\phi}),
\end{eqnarray} 
Eq.~\ref{Eq:g and h} results in
\begin{eqnarray}\label{phi and psi 1a}\
\Psi_k(\textbf{M}-\textbf{N})&=&|\lambda_k|\Phi_k\\
\Phi_k(\textbf{M}+\textbf{N})&=&|\lambda_k|\Psi_k
\label{phi and psi 1b}
\end{eqnarray} 
or equivalently
\begin{eqnarray}\label{phi and psi 2a}\
\Psi_k(\textbf{M}-\textbf{N})(\textbf{M}+\textbf{N})&=&|\lambda_k|^2\Psi_k\\
\Phi_k(\textbf{M}+\textbf{N})(\textbf{M}-\textbf{N})&=&|\lambda_k|^2\Phi_k
\label{phi and psi 2b}.
\end{eqnarray} 
where $(\Phi_k)_i=\phi_{ki}$, and $(\Psi_k)_i=\psi_{ki}$. Therefore $\Psi_k$ and $\lambda_k$ can be calculated by solving the eigenvalue equation (\ref{phi and psi 2a}), 
and for $\lambda_k\neq0$, $\Phi_k$ can be determined using (\ref{phi and psi 1a}). For $\lambda_k=0$, one should solve Eq. \ref{phi and psi 1b} directly to obtain $\Phi_k$.\\

Having obtained $\textbf{h}$ and $\textbf{g}$, one can calculate the correlation matrix $\textbf{G}$ for the full system defined as
\begin{eqnarray}\label{Gmat2}
G_{ij}&=&\langle (c_{i}^{\dagger}-c_{i})(c_{j}^{\dagger}+c_{j})\rangle
\end{eqnarray}
In terms of $\textbf{h}$ and $\textbf{g}$, $\textbf{G}$ can be also calculated as follows:
\begin{eqnarray}\label{Gmat7}
\textbf{G}= (\textbf{h}^{\dagger}-\textbf{g}^{\dagger})(\textbf{g}+\textbf{h}) .
\end{eqnarray}
In the following sections we use the above construction to diagonalize the $XY$ Hamiltonian.
\section{Diagonalization of the staggered $XY$ model}\label{SEC:solution}
In this section we present the details of diagonalization of the (modified) $XY$ Hamiltonian. The observable 
of interest here is the formation probability of the staggered pattern. The general scheme is to apply a projection transformation in such a way that this probability 
becomes an  EFP in a modified $XY$ Hamiltonian.\\

The formation probability for the staggered configuration $\left| \downarrow\uparrow\downarrow\uparrow...\right\rangle$ at zero temperature is:
\begin{widetext}
\begin{equation}
\begin{split}
p_{\text{stag}}(n)&=\left\langle 0\right| \left( \frac{1-\sigma_1^z}{2}\right) \left( \frac{1+\sigma_2^z}{2}\right) \left( \frac{1-\sigma_3^z}{2}\right) ...\left( \frac{1+(-1)^n\sigma_n^z}{2}\right) \left| 0\right\rangle \\
&=\left\langle 0\right|\left( \frac{1-\sigma_1^z}{2}\right) \sigma_2^x\left( \frac{1-\sigma_2^z}{2}\right)\sigma_2^x \left( \frac{1-\sigma_3^z}{2}\right) ...(\sigma_n^x)^{n+1}\left( \frac{1-\sigma_n^z}{2}\right)(\sigma_n^x)^{n+1}\left| 0\right\rangle\\
&=\left( \left\langle 0\right| \prod_{j=1}^{\text{int}(\frac{n}{2})}\sigma_{2j}^x\right) \prod_{j=1}^n\left( \frac{1-\sigma_j^z}{2}\right)\left( \prod_{j=1}^{\text{int}(\frac{n}{2})}\sigma_{2j}^x\left| 0\right\rangle\right)
\end{split}
\end{equation}
\end{widetext}
To go to the fermionic section, we use the JW transformation (Eq.~\ref{Eq:JW}). After applying $P_x=\prod_{j=1}^{\text{int}(\frac{n}{2})}\sigma_{2j}^x$, and also J-W transformation we obtain Eq.~\ref{Eq:modifiedXY}, from which we obtain:
\begin{equation}
\begin{split}
&H'_{XY}=\\
&\frac{1}{2}\sum_{l=1}^{L-1}\left[ J_n(l) \left( c_l^{\dagger}c_{l+1}-c_lc_{l+1}^{\dagger} \right) +J\gamma_n(l) \left( c_l^{\dagger}c_{l+1}^{\dagger}-c_lc_{l+1}\right)\right]\\
&-\frac{NJ}{2} \left( c_L^{\dagger}c_1+\gamma c_L^{\dagger}c_1^{\dagger}+H.C.\right)-\sum_{l=1}^Lh_n(l)c_l^{\dagger}c_l+cnst.
\end{split}
\end{equation}
where the constants were defined in Eq.~\ref{Eq:gamma-J}. Also note that $c_{L+1}=-Nc_1$, from which we see that for $N=-1$ one obtains:
\begin{equation}
\begin{split}
&H'_{XY}=\\
&\frac{1}{2}\sum_{l=1}^L\left[ J_n(l) \left( c_l^{\dagger}c_{l+1}-c_lc_{l+1}^{\dagger} \right) +J\gamma_n(l) \left( c_l^{\dagger}c_{l+1}^{\dagger}-c_lc_{l+1}\right)\right]\\
& -\sum_{l=1}^Lh_n(l)c_l^{\dagger}c_l+const.
\label{Eq:effectiveH}
\end{split}
\end{equation}
If we write the modified $XY$ Hamiltonian in the following form:
\begin{equation}
H'_{XY}=\sum_{i,j}\left[c_i^{\dagger}A_{i,j}c_j +\frac{1}{2}\left(c_i^{\dagger}B_{i,j}c_j^{\dagger}+ h.c.\right) \right]
\label{Eq:Ising_AB}
\end{equation}
then we have:
\begin{equation}
\begin{split}
&A_{ij}(i\leq n \ \text{and}\ j\leq n)=\left\lbrace \begin{matrix}
(-1)^ih & \text{if} \ \ i=j\\
\frac{1}{2}\gamma J & \text{if} \ \ i=j\pm 1
\end{matrix}\right. \\ 
&A_{ij}(i> n \ \text{or}\ j> n)=\left\lbrace \begin{matrix}
-h & \text{if} \ \ i=j\\
\frac{1}{2} J & \text{if} \ \ i=j\pm 1\\
\frac{1}{2} J  &  \text{if} \ \ i=1,j=L \ \text{or} \ \ i=L,j=1\\
0 & \text{otherwise}
\end{matrix}\right. 
\end{split}
\label{Eq:A}
\end{equation}
Also 
\begin{equation}
\begin{split}
&B_{ij}(i\leq n \ \text{and}\ j\leq n)=\left\lbrace \begin{matrix}
\frac{1}{2} J & \text{if} \ \ i=j+1\\
-\frac{1}{2} J & \text{if} \ \ i=j-1
\end{matrix}\right. \\ 
&B_{ij}(i> n \ \text{or}\ j> n)=\left\lbrace \begin{matrix}
\frac{1}{2} \gamma J & \text{if} \ \ i=j+1\\
-\frac{1}{2} \gamma J & \text{if} \ \ i=j-1\\
-\frac{1}{2} \gamma J  &  \text{if} \ \ i=1,j=L\\
\frac{1}{2} \gamma J  &  \text{if} \ \ i=L,j=1\\
0 & \text{otherwise}
\end{matrix}\right.
\end{split}
\label{Eq:B}
\end{equation}
Therefore one obtains the following form for $\textbf{C}\equiv(\textbf{A}-\textbf{B})(\textbf{A}+\textbf{B})$:
\begin{equation}
\begin{split}
&C_{ii}=h^2+\frac{1}{2}J^2(1+\gamma^2)\\
&C_{i,i+1}=\left\lbrace \begin{matrix}
(-1)^i hJ & \text{if} \ \ i\leq n\\
-hJ  &  \text{if} \ \ i> n
\end{matrix}\right. \\
&C_{i,i+2}=\left\lbrace \begin{matrix}
-\frac{1}{4}J^2(1-\gamma^2) & \text{if} \ \ i\leq n\\
\frac{1}{4}J^2(1-\gamma^2) & \text{if} \ \ i> n
\end{matrix}\right.
\end{split}
\end{equation}
Note that $\textbf{C}$ is symmetric, and also $C_{L-1,1}=C_{L,2}=C_{1,L-1}=C_{2,L}=\frac{1}{4}J^2(1-\gamma^2)$, and also $C_{1,L}=C_{L,1}=-hJ$. All other components of $\textbf{C}$ are zero. We use Eq.~\ref{Eq:pairfunction} to diagonalize this matrix and obtain $\Lambda_k$'s and also $\Psi$. We analyze two cases separately: $m\ll \frac{n}{2}$ (deep inside the staggered interval, or the DISI region), and $m\gg \frac{n}{2}$ (deep outside the staggered interval, or the DOSI region).\\
\\
\textbf{DISI} case:\\
\\
For the solution, we consider the trial function Eq.~\ref{Eq:pairfunction} with constants $a_1$, $a_2$, $a_3$ and $a_4$ coefficients to be determined. This function can be re-written in the following form:
\begin{equation}
\begin{split}
\psi(m)=\left\lbrace \begin{matrix}
 a_1\sin km +a_2\cos km & \ \  \text{for the odd sublattice} \\
 a_3\sin km +a_4\cos km & \ \  \text{for the even sublattice}
\end{matrix}\right. 
\end{split}
\end{equation}
In this case, applying Eq.~\ref{phi and psi 2a}, we end up with two set of equations (due to the bipartite nature of the lattice) to be solved (for DISI):
\begin{widetext}
\begin{equation}
\begin{split}
(1)\ \  &\sin km\left(-\frac{1}{2}a_1J^2(1-\gamma^2)\cos k+a_3hJ(\cos k-1)+a_4hJ\sin k  +a_1(h^2+\frac{1}{2}J^2(1+\gamma^2)-\Lambda_k^2)\right) \\
&+\cos km\left(-\frac{1}{2}a_2J^2(1-\gamma^2)\cos k-a_3hJ\sin k+a_4hJ(\cos k-1)  +a_2(h^2+\frac{1}{2}J^2(1+\gamma^2)-\Lambda_k^2)\right)=0\\
(2)\ \  & \ \  \sin km\left(-\frac{1}{2}a_3J^2(1-\gamma^2)\cos k+a_1hJ(\cos k-1)-a_2hJ\sin k  +a_3(h^2+\frac{1}{2}J^2(1+\gamma^2)-\Lambda_k^2)\right) \\
&+\cos km\left(-\frac{1}{2}a_4J^2(1-\gamma^2)\cos k+a_1hJ\sin k+a_2hJ(\cos k-1)  +a_4(h^2+\frac{1}{2}J^2(1+\gamma^2)-\Lambda_k^2)\right)=0
\end{split}
\end{equation}
\end{widetext}
Each component (the coefficients of $\sin km$ and $\cos km$) should be separately set to zero. Therefore, we obtain:
\begin{equation}
\begin{split}
\left\lbrace \begin{matrix}
a_1\zeta_k +a_3hJ(\cos k-1)+a_4hJ\sin k=0\\
a_2\zeta_k -a_3hJ\sin k +a_4hJ(\cos k-1)=0\\
a_3\zeta_k +a_1hJ(\cos k-1)-a_2hJ\sin k=0\\
a_4\zeta_k +a_1hJ\sin k +a_2hJ(\cos k-1)=0
\end{matrix}\right.
\end{split}
\end{equation}
where $\zeta_k\equiv -\frac{1}{2}J^2(1-\gamma^2)\cos k+h^2+\frac{1}{2}J^2(1+\gamma^2)-\Lambda_k^2$. In the matrix form, we have:
\begin{widetext}
\begin{equation}
\begin{split}
\left[  \begin{matrix}
\zeta_k & 0 & hJ(\cos k -1) & hJ\sin k\\
0 & \zeta_k & -hJ\sin k & hJ(\cos k-1)\\
hJ(\cos k-1) & -hJ\sin k & \zeta_k & 0 \\
hJ\sin k & hJ(\cos k-1) & 0 & \zeta_k
\end{matrix}\right] \left[  \begin{matrix}
a_1 \\
a_2 \\
a_3 \\
a_4
\end{matrix}\right]=0
\end{split}
\end{equation}
\end{widetext}
By setting the determinant to zero, we find that the eigenvalues should be of the following form:
\begin{equation}
\Lambda_k^2=\frac{1}{2}\left[2h^2+J^2(1+\gamma^2)-J^2(1-\gamma^2)\cos k\pm 4hJ\sin \frac{k}{2}  \right]
\label{Eq:deep1-energy}
\end{equation}
and the corresponding eigenvectors are:
\begin{equation}
\begin{split}
&\eta_1^-=\left( \begin{matrix}
-\cos \frac{k}{2} \\
\sin \frac{k}{2} \\
0 \\
1
\end{matrix}\right) \ , \ \eta_2^-=\left( \begin{matrix}
\sin \frac{k}{2} \\
\cos \frac{k}{2} \\
1 \\
0
\end{matrix}\right) \\ 
&\eta_1^+=\left( \begin{matrix}
\cos \frac{k}{2} \\
-\sin \frac{k}{2} \\
0 \\
1
\end{matrix}\right) \ , \ \eta_2^+=\left( \begin{matrix}
-\sin \frac{k}{2} \\
-\cos \frac{k}{2} \\
1 \\
0
\end{matrix}\right) 
\end{split}
\end{equation}
where the minus (plus) sign refers to the minus (plus) sign in the eigenvalues.\\
\\
\textbf{DOSI}\\
\\
Now let us work with DOSI following the same steps as the DISI case. Let us consider the coefficients to be $b_1$, $b_2$, $b_3$ and $b_4$. The equation governing $\Psi$ results in the following linear equations:
\begin{widetext}
\begin{equation}
\begin{split}
(1)\ \ &\sin km\left(\frac{1}{2}b_1J^2(1-\gamma^2)\cos k-b_3hJ(\cos k+1)-b_4hJ\sin k  +b_1(h^2+\frac{1}{2}J^2(1+\gamma^2)-\Lambda_k^2)\right) \\
&+\cos km\left(\frac{1}{2}b_2J^2(1-\gamma^2)\cos k+b_3hJ\sin k-b_4hJ(\cos k+1)  +b_2(h^2+\frac{1}{2}J^2(1+\gamma^2)-\Lambda_k^2)\right)=0\\
(2)\ \  &\sin km\left(\frac{1}{2}b_3J^2(1-\gamma^2)\cos k-b_1hJ(\cos k+1)+b_2hJ\sin k  +b_3(h^2+\frac{1}{2}J^2(1+\gamma^2)-\Lambda_k^2)\right) \\
&+\cos km\left(\frac{1}{2}b_4J^2(1-\gamma^2)\cos k-b_1hJ\sin k-b_2hJ(\cos k+1)  +b_4(h^2+\frac{1}{2}J^2(1+\gamma^2)-\Lambda_k^2)\right)=0
\end{split}
\end{equation}
\end{widetext}
resulting in:
\begin{equation}
\begin{split}
\left\lbrace \begin{matrix}
b_1\zeta'_k -b_3hJ(\cos k+1)-b_4hJ\sin k=0\\
b_2\zeta'_k +b_3hJ\sin k -b_4hJ(\cos k+1)=0\\
b_3\zeta'_k +b_1hJ(\cos k+1)+b_2hJ\sin k=0\\
b_4\zeta'_k -b_1hJ\sin k -b_2hJ(\cos k+1)=0
\end{matrix}\right. ,
\end{split}
\end{equation} 
or, in the matrix form:
\begin{widetext}
\begin{equation}
\begin{split}
\left[  \begin{matrix}
\zeta'_k & 0 & -hJ(\cos k +1) & -hJ\sin k\\
0 & \zeta'_k & hJ\sin k & -hJ(\cos k+1)\\
-hJ(\cos k+1) & hJ\sin k & \zeta'_k & 0 \\
-hJ\sin k & -hJ(\cos k+1) & 0 & \zeta'_k
\end{matrix}\right] \left[  \begin{matrix}
b_1 \\
b_2 \\
b_3 \\
b_4
\end{matrix}\right]=0
\end{split}
\end{equation}
\end{widetext}
where $\zeta'_k\equiv \frac{1}{2}J^2(1-\gamma^2)\cos k+h^2+\frac{1}{2}J^2(1+\gamma^2)-\Lambda_k^2$. The corresponding eigenvalues are:
\begin{equation}
\Lambda_k^2=\frac{1}{2}\left[2h^2+J^2(1+\gamma^2)+J^2(1-\gamma^2)\cos k\pm 4hJ\cos \frac{k}{2}  \right]
\label{Eq:deep2-energy}
\end{equation}
This form is just like the eigenvalues found for the DISI, with different sign for $J^2(1-\gamma^2)\cos k$. The corresponding eigenvectors are:
\begin{equation}
\begin{split}
&\eta_1^-=\left( \begin{matrix}
\sin \frac{k}{2} \\
\cos \frac{k}{2} \\
0 \\
1
\end{matrix}\right) \ , \ \eta_2^-=\left( \begin{matrix}
\cos \frac{k}{2} \\
-\sin \frac{k}{2} \\
1 \\
0
\end{matrix}\right) \\ 
&\eta_1^+=\left( \begin{matrix}
-\sin \frac{k}{2} \\
-\cos \frac{k}{2} \\
0 \\
1
\end{matrix}\right) \ , \ \eta_2^+=\left( \begin{matrix}
-\cos \frac{k}{2} \\
\sin \frac{k}{2} \\
1 \\
0
\end{matrix}\right) 
\end{split}
\end{equation}
where again the the minus (plus) sign refers to the minus (plus) sign in the eigenvalues. Since we pair the sites in the direct space, the size of the first Brillouin zone is doubled. If we use the natural change $k=2q$, then we obtain:
\begin{equation}
\begin{split}
\Lambda_q^2&=\frac{1}{2}\left[2h^2+J^2(1+\gamma^2)+J^2(1-\gamma^2)\cos 2q\pm 4hJ\cos q  \right]\\
& = (J\cos q\pm h)^2+J^2\gamma^2\sin^2q
\end{split}
\end{equation}
that is exactly the spectrum of the single-particle energies of the Fermions.\\

Summarizing, for DOSI, (using the above $\eta$'s) the full eigenvector is readily calculated to be:
\begin{equation}
\begin{split}
&\psi_1^-=\left\lbrace   \begin{matrix}
\cos k(m-\frac{1}{2}) \\
\cos km 
\end{matrix}\right. \ , \ \psi_2^-=\left\lbrace  \begin{matrix}
\sin k(m-\frac{1}{2}) \\
\sin km 
\end{matrix}\right. \\ 
&\psi_1^+=\left\lbrace \begin{matrix}
-\cos k(m-\frac{1}{2}) \\
\cos km 
\end{matrix}\right. \ , \ \psi_2^+=\left\lbrace \begin{matrix}
-\sin k(m-\frac{1}{2}) \\
\sin km 
\end{matrix}\right. 
\end{split}
\end{equation}
whereas for DISI, 
\begin{equation}
\begin{split}
&\psi_1^-=\left\lbrace   \begin{matrix}
-\sin k(m-\frac{1}{2}) \\
\cos km 
\end{matrix}\right. \ , \ \psi_2^-=\left\lbrace  \begin{matrix}
\cos k(m-\frac{1}{2}) \\
\sin km 
\end{matrix}\right. \\ 
&\psi_1^+=\left\lbrace \begin{matrix}
\sin k(m-\frac{1}{2}) \\
\cos km 
\end{matrix}\right. \ , \ \psi_2^+=\left\lbrace \begin{matrix}
-\cos k(m-\frac{1}{2}) \\
\sin km 
\end{matrix}\right. 
\end{split}
\end{equation}
Interestingly we see that the phase shift $k\rightarrow \pi-k$ relates the eigenvalues in DISI to the eigenvalues in DOSI. Since we require that these eigenvalues be equal for the case of line defect, we should apply this phase shift for one of the regions, e.g. DISI. Under this action, $\cos k(m-\frac{1}{2})\rightarrow -(-1)^m\sin k(m-\frac{1}{2})$, $\sin k(m-\frac{1}{2})\rightarrow -(-1)^m\cos k(m-\frac{1}{2})$, $\cos km\rightarrow (-1)^m\cos km$ and $\sin km\rightarrow -(-1)^m\sin km$. Therefore, for the DISI:
\begin{equation}
\begin{split}
&\psi_1^-\rightarrow\left\lbrace   \begin{matrix}
-(-1)^{m+1}\cos k(m-\frac{1}{2}) \\
(-1)^m\cos km 
\end{matrix}\right. = (-1)^m\times\left\lbrace   \begin{matrix}
\cos k(m-\frac{1}{2}) \\
\cos km 
\end{matrix}\right.\\
&\psi_2^-\rightarrow\left\lbrace  \begin{matrix}
(-1)^{m+1}\sin k(m-\frac{1}{2}) \\
(-1)^{m+1}\sin km 
\end{matrix}\right. \rightarrow (-1)^m\times\left\lbrace  \begin{matrix}
\sin k(m-\frac{1}{2}) \\
\sin km 
\end{matrix}\right.\\
&\psi_1^+\rightarrow\left\lbrace \begin{matrix}
(-1)^{m+1}\cos k(m-\frac{1}{2}) \\
(-1)^m\cos km 
\end{matrix}\right.=(-1)^m\times\left\lbrace \begin{matrix}
-\cos k(m-\frac{1}{2}) \\
\cos km 
\end{matrix}\right. \\ 
&\psi_2^+\rightarrow\left\lbrace \begin{matrix}
-(-1)^{m+1}\sin k(m-\frac{1}{2}) \\
(-1)^{m+1}\sin km 
\end{matrix}\right.=(-1)^m\times\left\lbrace \begin{matrix}
\sin k(m-\frac{1}{2}) \\
-\sin km 
\end{matrix}\right. 
\end{split}
\end{equation}
showing that $\psi_{DISR}=(-1)^m\psi_{DOSR}$. \\

Now let us consider the chain as a whole, for which the boundary conditions at $j=n$ should be worked out. To this end, we mix the two solutions obtained above. Based on the above findings, the following trial function is considered ($m_0\equiv \left[\frac{n}{2} \right] $):
\begin{equation}
\begin{split}
&\psi(m\leq m_0)=(-1)^m\times\left[ 
\begin{matrix}
a\cos k(m-\frac{1}{2})+b\sin k(m-\frac{1}{2})\\
a \cos km +b\sin km
\end{matrix}\right. \\
&\psi(m> m_0)=\times\left[ 
\begin{matrix}
c\cos k(m-\frac{1}{2})+d\sin k(m-\frac{1}{2})\\
c \cos km +d\sin km
\end{matrix}\right.
\end{split}
\label{Eq:trial_wave}
\end{equation}
with undetermined constants $a$, $b$, $c$, and $d$, to be found by applying the boundary conditions. To facilitate this calculation let us consider $\gamma=1$ (pure Ising model). We have seen that the solution for the Ising model is also valid for generic $\gamma$. For $\gamma=1$ we have four independent equations at the boundaries:
\begin{widetext}
\begin{equation}
\begin{split}
&(a\cos km_0+b\sin km_0)-(c\cos k(m_0+1)+d\sin k(m_0+1))+2\cos\frac{k}{2}(c\cos k(m_0+\frac{1}{2})+d\sin k(m_0+\frac{1}{2}))=0\\
&-(c\cos\frac{kL}{2}+d\sin\frac{kL}{2})+(a\cos k+b\sin k)+2\cos\frac{k}{2}(-a\cos\frac{k}{2}-b\sin\frac{k}{2})=0\\
& a\cos\frac{k}{2}+b\sin\frac{k}{2}-(c\cos\frac{k(L-1)}{2}+d\sin\frac{k(L-1)}{2})+2\cos\frac{k}{2}(c\cos\frac{kL}{2}+d\sin\frac{kL}{2})=0\\
&-(a\cos k(m_0-\frac{1}{2})+b\sin k(m_0-\frac{1}{2}))+(c\cos k(m_0+\frac{1}{2})+d\sin k(m_0+\frac{1}{2}))+2\cos \frac{k}{2}(a\cos km_0+b\sin km_0)=0
\end{split}
\end{equation}
\end{widetext}
These equations can be written in the matrix form:
\begin{equation}
\begin{split}
\left[ \begin{matrix}
a_{11} & a_{12} & a_{13} & a_{14} \\
a_{21} & a_{22} & a_{23} & a_{24} \\
a_{31} & a_{32} & a_{33} & a_{34} \\
a_{41} & a_{42} & a_{43} & a_{44} 
\end{matrix}\right]\left[ \begin{matrix}
a \\
b \\
c \\
d
\end{matrix}\right] =0
\end{split}
\end{equation}
with the elements: 
\begin{equation}
\begin{split}
&a_{11}=\cos km_0,a_{12}=\sin km_0,a_{13}=\cos k m_0,a_{14}=\sin km_0, \\
&a_{21}=-1,\ a_{22}=0,\ a_{23}=-\cos\frac{kL}{2},\ a_{24}=-\sin\frac{kL}{2},\\
&a_{31}=\cos\frac{k}{2},\ a_{32}=\sin\frac{k}{2}, a_{33}=\cos\frac{k(L+1)}{2},\\
&a_{34}=\sin\frac{k(L+1)}{2},a_{41}=\cos k(m_0+\frac{1}{2}),a_{42}=\sin k(m_0+\frac{1}{2})\\
&a_{43}=\cos k(m_0+\frac{1}{2}),\ a_{44}=\sin k(m_0+\frac{1}{2})
\end{split}
\end{equation}
The determinant of this matrix is $-4\sin^2\frac{k}{2}\sin^2\frac{kL}{2}$, the zeros of which take place at $k_s=\frac{4\pi s}{L}$ in accordance with the free case (the case with no staggered interval involved). This is expected since the single particle energy spectrum of the original $XY$ model should not change under the action of the unitary transformation $P_x$. For $\sin k_sm_0 \neq 0$ the solution for $a$, $b$, $c$ and $d$ is:
\begin{equation}
a=-\cot k_sm_0, \ \ \ b=-1, \ \ \ c=\cot\left(k_sm_0\right), \ \ \ d=1
\end{equation}
This results in
\begin{equation}
\begin{split}
&\psi(m\leq m_0)=-(-)^m\psi_0\times\left\lbrace 
\begin{matrix}
\cos k_s\left[m-m_0-\frac{1}{2}\right]\\
\\
\cos k_s\left[m-m_0\right]
\end{matrix}\right. \\
&\psi(m> m_0)=\psi_0\left\lbrace  
\begin{matrix}
\cos k_s\left[m-m_0-\frac{1}{2}\right]\\
\\
\cos k_s\left[m-m_0\right]
\end{matrix}\right.
\end{split}
\end{equation}
where $\psi_0$ is the normalization factor, which is shown to be $\sqrt{\frac{2}{L}}$. In terms of $j=2m_j$ and $q_s=\frac{k_s}{2}$ we find that:
\begin{equation}
\begin{split}
\psi(j)=\sqrt{\frac{2}{L}}\left\lbrace \begin{matrix}
-(-1)^{m_j}
\cos q_s\left[j-n\right] & j\leq n\\
\cos q_s\left[j-n\right] & j> n
\end{matrix}\right. 
\end{split}
\label{Eq:psi}
\end{equation}
Also the other choice for $a$, $b$, $c$ and $d$ is
\begin{equation}
\begin{split}
& a=\tan k_sm_0, \ \ \ b=-1, \ \ \ c=-\tan k_sm_0,\ \ \ d=1
\end{split}
\end{equation}
which is equivalent to $\cos \leftrightarrow \sin$. Although Eq.~\ref{Eq:psi} was obtained for $\gamma=1$, it is a general result for the staggered line defect, and is valid for generic $\gamma$.

\section{Correlation functions}\label{SEC:corr_func}

Here we present the details of calculation of the correlation functions. In the previous appendix we showed that the eigenvector of the matrix $\textbf{C}$ for the $XY$ model in the general form is:
\begin{equation}
\psi_{XY}=\sqrt{\frac{2}{L}}\left\lbrace \begin{matrix}
-(-)^{m_j}\cos q_s (j-n) & j\leq n\\
\cos q_s (j-n) & j> n
\end{matrix} \right. 
\end{equation}
Therefore using the relation~\ref{phi and psi 1b} we obtain the general form of $\phi_{si}$ for $\Lambda_k \ne 0$ (generic $\gamma$):
\begin{widetext}
\begin{equation}
\begin{split}
\phi_{XY}(j\leq n)=&-\Lambda_{q_s}^{-1}\sqrt{\frac{2}{L}}\left[(-)^{m_j+a_j}h\cos q_s(j-n)\right. \\
&\left. +\frac{-1+\gamma}{2}(-)^{m_{j-1}}\cos q_s(j-n-1)+\frac{1+\gamma}{2}(-)^{m_{j+1}}\cos q_s(j-n+1) \right] \\
\phi_{XY}(j= n+1)=&-\Lambda_{q_s}^{-1}\sqrt{\frac{2}{L}}\left[h\cos q_s(j-n)\right. \\
&\left. \frac{-1+\gamma}{2}(-)^{m_{j-1}}\cos q_s(j-n-1)-\frac{1+\gamma}{2}\cos q_s(j-n+1) \right] \\
\phi_{XY}(j> n+1)=&-\Lambda_{q_s}^{-1}\sqrt{\frac{2}{L}}\left[h\cos q_s(j-n)\right. \\
&\left. +\frac{-1+\gamma}{2}\cos q_s(j-n-1)-\frac{1+\gamma}{2}\cos q_s(j-n+1) \right] 
\end{split}
\end{equation}
\end{widetext}
We should have $\sin \leftrightarrow \cos$ as we go from positive $q_s$'s to negative ones. Let us work with $f_{AB}^{i,j}=\sum_{s=-\frac{L}{2}}^{L/2-1}A_{si}B_{sj}$, in which $A,B=\psi, \phi$. The importance of these functions can be understood noting that:
\begin{equation}
\begin{split}
&\left\langle c_ic_j^{\dagger} \right\rangle =\frac{1}{4}\left[ f_{\psi\psi}+f_{\psi\phi}+f_{\phi\psi}+f_{\phi\phi}\right],\\ 
&\left\langle c_ic_j \right\rangle =\frac{1}{4}\left[f_{\psi\psi}-f_{\psi\phi}+f_{\phi\psi}-f_{\phi\phi}\right], \\
&\left\langle c_i^{\dagger}c_j^{\dagger} \right\rangle=\frac{1}{4}\left[f_{\psi\psi}+f_{\psi\phi}-f_{\phi\psi}-f_{\phi\phi}\right], \\ 
&\left\langle c_i^{\dagger}c_j \right\rangle =\frac{1}{4}\left[f_{\psi\psi}-f_{\psi\phi}-f_{\phi\psi}+f_{\phi\phi}\right],
\end{split}
\label{Eq:correlations}
\end{equation}
One can easily show that always (irrespective to the amount of $i$ and $j$ being inside or outside the staggered interval) $f_{\psi\psi}^{i,j}=f_{\phi\phi}^{i,j}=\delta_{ij}$, and $\delta_{ij}$ is the Kronecker delta.\\

Having $\Psi$ and $\Phi$ in hand, one can directly calculate $f_{\psi\psi}$, $f_{\phi\phi}$, $f_{\psi\phi}$, and $f_{\phi\psi}$. We immediately obtain that $f_{\psi\psi}=\delta_{ij}$ as expected. In the following we prove also that $f_{\phi\phi}=\delta_{ij}$. Let us consider $i\leq n$ and $j\leq n$. Then we have:
\begin{widetext}
\begin{equation}
\begin{split}
f_{\phi\phi}=&\sum_s\phi_{si}\phi_{sj}=\frac{1}{L}\sum_s\left\lbrace (-)^{m_i+a_i-m_j-a_j}h^2+\left(\frac{-1+\gamma}{2} \right)^2(-)^{m_{j-1}-m_{i-1}}+ \left(\frac{1+\gamma}{2} \right)^2(-)^{m_{j+1}-m_{i+1}}\right\rbrace \cos q_s(j-i) \\
&+\frac{1}{L}\sum_s\left\lbrace (-)^{m_i+a_i-m_{j-1}}h\left(\frac{-1+\gamma}{2} \right)\cos q_s(j-i-1)+(-)^{m_i+a_i-m_{j+1}}h\left(\frac{1+\gamma}{2} \right)\cos q_s(j-i+1)\right\rbrace \\
&+\frac{1}{L}\sum_s\left\lbrace (-)^{m_{i-1}-m_j-a_j}h\left(\frac{-1+\gamma}{2} \right)\cos q_s(j-i+1)+(-)^{m_{i-1}-m_{j+1}}\left(\frac{-1+\gamma}{2} \right)\left(\frac{1+\gamma}{2} \right)\cos q_s(j-i+2)\right\rbrace \\
&+\frac{1}{L}\sum_s\left\lbrace (-)^{m_{i+1}-m_j-a_j}h\left(\frac{1+\gamma}{2} \right)\cos q_s(j-i-1)+(-)^{m_{i+1}-m_{j-1}}\left(\frac{-1+\gamma}{2} \right)\left(\frac{1+\gamma}{2} \right)\cos q_s(j-i-2)\right\rbrace
\end{split}
\end{equation}
\end{widetext}
Noting that $(-)^{m_{i-1}}=(-)^{m_i+a_i}=-(-)^{m_{i+1}}$, we obtain:
\begin{widetext}
\begin{equation}
\begin{split}
f_{\phi\phi}=&(-)^{m_{j+1}-m_{i+1}}\frac{1}{L}\sum_s\Lambda_s^{-2}\left(h^2+\frac{1}{2}(1+\gamma^2) \right)\cos q_s(j-i)\\
& +(-)^{m_{j+1}-m_{i+1}}h\left(\frac{-1+\gamma}{2}-\frac{1+\gamma}{2} \right) \frac{1}{L}\sum_s\Lambda_s^{-2}\cos q_s(j-i-1)\\
& +(-)^{m_{j+1}-m_{i+1}}h\left(\frac{-1+\gamma}{2}-\frac{1+\gamma}{2} \right) \frac{1}{L}\sum_s\Lambda_s^{-2}\cos q_s(j-i+1)\\
& -(-)^{m_{j+1}-m_{i+1}}\left(\frac{\gamma^2-1}{4}\right) \frac{1}{L}\sum_s\Lambda_s^{-2}\left[ \cos q_s(j-i-2)+\cos q_s(j-i+2)\right]\\
&=(-)^{m_{j+1}-m_{i+1}}\frac{1}{L}\sum_s\Lambda_s^{-2}\left[h^2+\frac{1}{2}\gamma^2\left(1-\cos 2q_s \right)+\frac{1}{2}\left(1+\cos 2q_s \right)-2h\cos q_s \right] \cos q_s(j-i)\\
&=(-)^{m_{j+1}-m_{i+1}}\frac{1}{L}\sum_s\Lambda_s^{-2}\left[h^2+\gamma^2\sin^2q_s+\cos^2q_s-2h\cos q_s \right] \cos q_s(j-i)\\
&=(-)^{m_{j+1}-m_{i+1}}\frac{1}{L}\sum_s\cos q_s(j-i)=\delta_{ij}=f_{\psi\psi}\\
\end{split}
\end{equation}
\end{widetext}
leading to $\left\lbrace c_i^{\dagger},c_j\right\rbrace=\frac{1}{2}\left(f_{\psi\psi}+f_{\phi\phi}\right)=\delta_{ij} $ as expected. Let us next calculate $f_{\psi\phi}$ and $f_{\phi\psi}$:
\begin{widetext}
\begin{equation}
\begin{split}
f_{\psi\phi}=&(-)^{m_j+a_j-m_i}h\frac{1}{L}\sum_s\Lambda_s^{-1}\cos q_s(j-i)\\
&+(-)^{m_{j-1}-m_i}\left(\frac{-1+\gamma}{2} \right) \frac{1}{L}\sum_s\Lambda_s^{-1}\cos q_s(j-i-1)\\
&+(-)^{m_{j+1}-m_i}\left(\frac{1+\gamma}{2} \right) \frac{1}{L}\sum_s\Lambda_s^{-1}\cos q_s(j-i+1)\\
&=(-)^{m_{j+1}-m_i}\frac{1}{L}\sum_s\Lambda_s^{-1}\left[ -h\cos q_s(j-i)+\left(\frac{1-\gamma}{2} \right)\cos q_s(j-i-1)+\left(\frac{1+\gamma}{2} \right)\cos q_s(j-i+1)\right]\\
&=(-)^{m_{j+1}-m_i}\frac{1}{L}\sum_s\Lambda_s^{-1}\left[-h\cos q_s(j-i)+\cos q_s\cos q_s(j-i)-\gamma\sin q_s\sin q_s(j-i)\right] 
\end{split}
\end{equation}
\end{widetext}
\begin{widetext}
\begin{equation}
\begin{split}
f_{\phi\psi}=&(-)^{m_i+a_i-m_j}h\frac{1}{L}\sum_s\Lambda_s^{-1}\cos q_s(j-i)\\
&+(-)^{m_{i-1}-m_j}\left(\frac{-1+\gamma}{2} \right) \frac{1}{L}\sum_s\Lambda_s^{-1}\cos q_s(j-i+1)\\
&+(-)^{m_{i+1}-m_j}\left(\frac{1+\gamma}{2} \right) \frac{1}{L}\sum_s\Lambda_s^{-1}\cos q_s(j-i-1)\\
&=(-)^{m_{i+1}-m_j}\frac{1}{L}\sum_s\Lambda_s^{-1}\left[ -h\cos q_s(j-i)+\left(\frac{1-\gamma}{2} \right)\cos q_s(j-i+1)+\left(\frac{1+\gamma}{2} \right)\cos q_s(j-i-1)\right]\\
&=(-)^{m_i-m_j+a_i}\frac{1}{L}\sum_s\Lambda_s^{-1}\left[ h\cos q_s(j-i)-\cos q_s\cos q_s(j-i)-\gamma\sin q_s\sin q_s(j-i)\right]\\
&=(-)^{m_j-m_i+a_j}\frac{1}{L}\sum_s\Lambda_s^{-1}(-)^{a_i-a_j}\left[ h\cos q_s(j-i)-\cos q_s\cos q_s(j-i)-\gamma\sin q_s\sin q_s(j-i)\right]\\
&=(-)^{m_{j+1}-m_i}\frac{1}{L}\sum_s\Lambda_s^{-1}(-)^{a_i-a_j}\left[-h\cos q_s(j-i)+\cos q_s\cos q_s(j-i)+\gamma\sin q_s\sin q_s(j-i)\right]
\end{split}
\end{equation}
\end{widetext}
Therefore 
\begin{equation}
\begin{split}
\frac{1}{4}\left( f_{\psi\phi}\pm f_{\phi\psi}\right) &=\frac{1}{2}(-)^{m_{j+1}-m_i}\sum_s\Lambda_s^{-1}\left\lbrace \left[\frac{1\pm(-)^{a_i-a_j}}{2} \right]\times \right. \\
&\left. \left(-h\cos q_s(j-i)+\cos q_s\cos q_s(j-i)\right) \right.\\
& \left. -\gamma\left[\frac{1\mp (-)^{a_i-a_j}}{2}\right]\sin q_s\sin q_s(j-i)\right\rbrace \\
&=\chi_{ij}^{\pm}\sigma_1(i,j)+\chi_{ij}^{\mp}\sigma_2(i,j)
\end{split}
\end{equation}
where $\chi_{ij}^{+}=\left[\frac{1+(-)^{a_i-a_j}}{2} \right](-)^{m_{j+1}-m_i}$ and $\chi_{ij}^-=\left[\frac{1-(-)^{a_i-a_j}}{2} \right](-)^{m_{j+1}-m_i}$, and also 
\begin{equation}
\begin{split}
\sigma_1(j,k)&=\frac{1}{2L}\sum_{s=-L/2}^{L/2-1}\cos q_s(k-j)\left( \frac{-h+\cos q_s}{\Lambda_s}\right)\\
&=\frac{1}{2L}\sum_{s=-L/2}^{L/2-1}\left( \frac{-h+\cos q_s}{\Lambda_s}\right)e^{-iq_s(k-j)}\\
\sigma_2(j,k)&=\frac{1}{2L}\sum_{s=-L/2}^{L/2-1}\sin q_s(k-j)\left( \frac{\gamma\sin q_s}{\Lambda_s}\right)\\
&=\frac{i}{2L}\sum_{s=-L/2}^{L/2-1}\left( \frac{\gamma\sin q_s}{\Lambda_s}\right)e^{-iq_s(k-j)}
\end{split}
\label{Eq:sigmas}
\end{equation} 
where we have used the symmetry considerations to add extra zero contributions. Therefore if $i$ and $j$ belong to the same sublattice, then $\chi_{ij}^+=(-)^{m_{j+1}-m_i}$ and $\chi_{ij}^-=0$, so that:
\begin{equation}
\begin{split}
&\Sigma_1(i,j)\equiv\frac{1}{4}\left( f_{\psi\phi}+ f_{\phi\psi}\right)=(-)^{m_{j+1}-m_i}\sigma_1(j-i), \\
&\Sigma_2(i,j)\equiv\frac{1}{4}\left( f_{\psi\phi}- f_{\phi\psi}\right)=(-)^{m_{j+1}-m_i}\sigma_2(j-i)
\end{split}
\end{equation}
Also if they belong to the different sublattices, then $\chi_{ij}^-=(-)^{m_{j+1}-m_i}$ and $\chi_{ij}^+=0$, and therefore we find that:
\begin{equation}
\begin{split}
&\Sigma_1(i,j)=(-)^{m_{j+1}-m_i}\sigma_2(j-i), \\
&\Sigma_2(i,j)=(-)^{m_{j+1}-m_i}\sigma_1(j-i)
\end{split}
\end{equation}
The correlation functions now can be determined explicitly. Using Eq.~\ref{Eq:correlations} we find:
\begin{equation}
\begin{split}
\left\langle c_ic_j^{\dagger} \right\rangle &=\frac{1}{2}\delta_{ij}+\Sigma_1(i,j)\\
&=\frac{1}{2}\delta_{ij}+\chi_{ij}^+\sigma_1(i,j)+\chi_{ij}^-\sigma_2(i,j)\\
\left\langle c_ic_j \right\rangle &=\Sigma_2(i,j)\\
&=\chi_{ij}^+\sigma_2(i,j)+\chi_{ij}^-\sigma_1(i,j)\\
\left\langle c_i^{\dagger}c_j^{\dagger} \right\rangle &=-\chi^+_{ij}\sigma_2(i,j)-\chi_{ij}^-\sigma_1(i,j)=-\left\langle c_ic_j \right\rangle\\
\left\langle c_i^{\dagger}c_j \right\rangle &=\frac{1}{2}\delta_{ij}-\chi_{ij}^+\sigma_1(i,j)-\chi_{ij}^-\sigma_2(i,j)=\delta_{ij}-\left\langle c_ic_j^{\dagger} \right\rangle
\end{split}
\end{equation}
Therefore, noting that $s_{ij}=\left\langle c_ic_j^{\dagger} \right\rangle - \left\langle c_ic_j \right\rangle $, we see:
\begin{equation}
s_{ij}=\frac{1}{2}\delta_{ij}+\frac{1}{2}(\chi_{ij}^+-\chi_{ij}^-)\sigma(i,j)
\end{equation}
where we have defined:
\begin{equation}
\begin{split}
\sigma(j,k)&\equiv 2\left( \sigma_1(j,k)-\sigma_2(j,k)\right) \\
&=\frac{1}{L}\sum_{s=-L/2}^{L/2-1}\left( \frac{\cos q_s-h-i\gamma\sin q_s}{\Lambda_s}\right)e^{iq_s(j-k)}
\end{split}
\end{equation}
which, in the $L\rightarrow \infty$ limit becomes
\begin{equation}
\begin{split}
&\sigma(j,k)=\int \frac{\text{d}q}{2\pi} \sigma(q)e^{iq(j-k)}\\
& \sigma(q)\equiv \frac{\cos q-h-i\gamma\sin q}{\Lambda_q}
\end{split}
\end{equation}
Noting also that:
\begin{equation}
\begin{split}
(-)^{m_{j+1}-m_i}&=\left\lbrace \begin{matrix}
(-)^{\frac{i-j-1}{2}} & \text{if} \ i, \ j\in \ (\text{different sublattices})\\
-(-)^{\frac{j+i}{2}} & \text{if} \ i,j\in \ (\text{same sublattice})
\end{matrix}\right.\\
&=-|\cos \frac{\pi}{2}(j-i)|(-)^{\frac{i+j}{2}}+|\sin \frac{\pi}{2}(j-i)| (-)^{\frac{i-j-1}{2}},
\end{split}
\end{equation}
one can easily verify that:
\begin{equation}
\begin{split}
s_{ij}=\frac{1}{2}\delta_{ij}+&\frac{1}{2}\cos\pi(j-i)\left\lbrace -(-)^{\frac{i+j}{2}}\left| \cos \frac{\pi}{2}(j-i)\right|\right. \\
& \left. +(-)^{\frac{i-j-1}{2}}\left| \sin \frac{\pi}{2}(j-i)\right| \right\rbrace \sigma(j-i)
\end{split}
\end{equation}

\bibliography{LineDefectEFPRef}




\end{document}